\pdfoutput=1
\documentclass{aa}  
\usepackage[varg]{txfonts}
\usepackage[T1]{fontenc}
\usepackage[utf8]{inputenc}
\usepackage{lmodern}
\usepackage{newunicodechar}
\usepackage[title]{appendix}

\usepackage[fleqn]{amsmath}
\usepackage{amsmath}
\usepackage{amstext}
\usepackage{mathtools}
\usepackage{amssymb}
\usepackage{xspace}
\usepackage{lipsum}
\usepackage{caption}
\usepackage{xcolor}
\usepackage{ragged2e}
\usepackage{graphicx}
\usepackage{graphicx,subfigure}
\usepackage{algorithm}
\usepackage[noend]{algpseudocode}
\usepackage[algo2e, algoruled, linesnumbered]{algorithm2e}% http://ctan.org/pkg/algorithm2e
\usepackage{algorithm2e}

\makeatletter
\newenvironment{breakablealgorithm}
  {% \begin{breakablealgorithm}
   \begin{center}
     \refstepcounter{algorithm}% New algorithm
     \hrule height.8pt depth0pt \kern2pt% \@fs@pre for \@fs@ruled
     \renewcommand{\caption}[2][\relax]{% Make a new \caption
       {\raggedright\textbf{\fname@algorithm~\thealgorithm} ##2\par}%
       \ifx\relax##1\relax % #1 is \relax
         \addcontentsline{loa}{algorithm}{\protect\numberline{\thealgorithm}##2}%
       \else % #1 is not \relax
         \addcontentsline{loa}{algorithm}{\protect\numberline{\thealgorithm}##1}%
       \fi
       \kern2pt\hrule\kern2pt
     }
  }{% \end{breakablealgorithm}
     \kern2pt\hrule\relax% \@fs@post for \@fs@ruled
   \end{center}
  }
\usepackage{graphicx}
\usepackage{txfonts}
\makeatother

\usepackage{natbib,twoopt}
\usepackage[breaklinks=true]{hyperref} %% to avoid \citeads line fills
\bibpunct{(}{)}{;}{a}{}{,}             %% natbib format for A&A and ApJ
\makeatletter
  \newcommandtwoopt{\citeads}[3][][]{\href{http://adsabs.harvard.edu/abs/#3}%
    {\def\hyper@linkstart##1##2{}%
     \let\hyper@linkend\@empty\citealp[#1][#2]{#3}}}
  \newcommandtwoopt{\citepads}[3][][]{\href{http://adsabs.harvard.edu/abs/#3}%
    {\def\hyper@linkstart##1##2{}%
     \let\hyper@linkend\@empty\citep[#1][#2]{#3}}}
  \newcommandtwoopt{\citetads}[3][][]{\href{http://adsabs.harvard.edu/abs/#3}%
    {\def\hyper@linkstart##1##2{}%
     \let\hyper@linkend\@empty\citet[#1][#2]{#3}}}
  \newcommandtwoopt{\citeyearads}[3][][]%
    {\href{http://adsabs.harvard.edu/abs/#3}
    {\def\hyper@linkstart##1##2{}%
     \let\hyper@linkend\@empty\citeyear[#1][#2]{#3}}}
\makeatother
\usepackage{hyperref}
\hypersetup{colorlinks=true,citecolor=blue,linkcolor=blue,linktocpage=true}
\begin{document}

\title{Reconstruction of weak lensing mass maps for non-Gaussian studies in the celestial sphere }
\titlerunning{Reconstruction of Weak Lensing mass maps in the celestial sphere}
% Authors names and emails
%\author{Vanshika Kansal \thanks{vanshikakansal@gmail.com}}
\author{Vanshika Kansal}

%\authorrunning{V. Kansal et al.}
\authorrunning{V. Kansal}

\institute{AIM, CEA, CNRS, Université Paris-Saclay, Université de Paris, F-91191 Gif-sur-Yvette, France \\
              \email{vanshikakansal@gmail.com}
}

   %\date{Received September 15, 1996; accepted March 16, 1997}

% \abstract{}{}{}{}{} 
% 5 {} token are mandatory
 
  \abstract
{
We present a novel method for reconstructing weak lensing mass or convergence maps as a probe to study non-Gaussianities in the cosmic density field. While previous surveys have relied on a flat-sky approximation, forthcoming Stage IV surveys will cover such large areas with a large field of view (FOV) to motivate mass reconstruction on the sphere. Here, we present an improved Kaiser-Squires (KS+) mass inversion method using a HEALPix pixelisation of the sphere while controlling systematic effects. As in the KS+ methodology, the convergence maps were reconstructed without noise regularisation to preserve the information content and allow for non-Gaussian studies. The results of this new method were compared with those of the Kaiser-Squires (KS) estimator implemented on the curved sky using high-resolution realistic N-body simulations. The quality of the method was evaluated by estimating the two-point correlation functions, third- and fourth-order moments, and peak counts of the reconstructed convergence maps. The effects of masking, sampling, and noise were tested. We also examined the systematic errors introduced by the flat-sky approximation. We show that the improved Kaiser-Squires on the sphere (SKS+) method systematically improves inferred correlation errors by $\sim$10 times and provides on average a 20–30 \% better maximum signal-to-noise peak estimation compared to Kaiser-Squires on the sphere (SKS). We also show that the SKS+ method is nearly unbiased and reduces errors by a factor of about 2 and 4 in the third- and fourth-order moments, respectively.
Finally, we show how the reconstruction of the convergence field directly on the celestial sphere eliminates the projection effects and allows the exclusion or consideration of a specific region of the sphere in the processing.
}


   \maketitle
%
%________________________________________________________________

\section{Introduction}

%__________________________________________________________________

Gravitational lensing is the deflection of light rays from distant sources by massive objects along the line of sight \citep[see, for example,][]{Bartelmann_2001}. The apparent shape of the source becomes distorted when the light ray is coherently distorted by the gravitational field of the matter acting as the lens.
In a weak lensing regime, where deflection angles are small, distortions in the image can be divided into two parts: shear, which distorts the shape of the source, and convergence, which isotropically magnifies or demagnifies it.
The measurement of these deformations offers an absolute gravitational probe for the total distribution of matter in the Universe (together with dark matter) and can be used to statistically infer cosmological information. 
Until now, cosmological analyses based on weak lensing mainly relied on two-point estimators of the shear field \citep[e.g.][]{Schneider_2002_2PCF}.
\justify
The most common approach is to compute a two-point shear correlation function from observational data \citep{Martin_2013, Troxel_2018, Hamana_2020, Hikage_2019, Asgari_2021, Secco_2022} or the shear power spectrum \citep{Hikage_2011, Heymans_2012, Lin_2012, Kohlinger_2017, Hikage_2019, Camacho_2021}, and compare it with theoretical expectations.
However, recently, cosmologists have become increasingly interested in extracting information from higher-order statistics such as peaks \citep[e.g.][]{ Hilbert_2012, Lin_2015, Martinet_2018, Peel_2018, Ajani_2020, Martinet_2021B, Martinet_2021A, Emma_2022}, higher-order moments \citep[e.g.][]{VanWaerbeke_2013, Petri_2015, Vicinanza_2016, Chang_2018, Peel_2018, Vicinanza_2018}, and Minkowski functionals \citep[e.g.][]{Kratochvil_2012, Petri_2015, Vicinanza_2019, Parroni_2020, Zurcher_2021}.
Although, theoretical predictions are more uncertain from these higher-order estimators. They are able to probe the non-Gaussian structure of the dark matter distribution and have proven to yield additional constraints \citep[e.g.][]{Liu_2015, Liu_2015B, Martinet_2018, Harnois_2021}.
These non-Gaussian estimators are typically calculated from the convergence field beyond what can be obtained from the shear field, as the convergence maps are rich in information about the interaction between galaxies and clusters. However, the convergence is not directly observable and must be reconstructed from shear measurements. This has motivated research into optimal mass-mapping techniques. \\

\citet{KS_1993} showed how to use weak lensing observable, shear measurements, to produce weak lensing convergence maps. This method is based on performing a direct Fourier inversion of the equations that relate the observed shear field to the convergence field, which is a scaled version of the integrated mass distribution.
Indeed, the Kaiser-Squires (KS) method has been used to recover mass maps from a variety of recent weak lensing surveys, the Canada-France-Hawaii Telescope Lensing Survey \citep[CFHTLenS\footnote{\url{https://www.cfhtlens.org}};][]{Heymans_2012} and the Dark Energy Survey \citep[DES\footnote{\url{https://www.darkenergysurvey.org}};][]{Flaugher_2015} Science Verification (SV) data \citep[respectively,][]{Massey_2007, VanWaerbeke_2013, Chang_2015}. An important limitation of this approach is that it does not take into account noise (noise growth on small scales), mask, or boundary effects (in practice, the resultant mass map is smoothed to mitigate noise). \\

It should be noted that several alternative mass map reconstruction methods have been proposed to address the limitations of the KS method. Some of these proposed techniques include Wiener filtering \citep{Jeffrey_2018}, sparsity priors \citep{Pires_2020}, and others \citep{Starck_2021, Jeffrey_2021}. One key assumption of these ‘planar’ mass-mapping techniques is that a small field of view of the celestial sphere can be well approximated as a tangential planar projection onto the celestial sphere. This assumption is colloquially referred to as the flat-sky approximation. For small-field surveys, this approximation is typically justified. However, such an assumption will not be appropriate for ongoing -- the Dark Energy Survey \citep[]{Flaugher_2015} and the Kilo Degree Survey \citep[KiDS\footnote{\url{http://www.astro-wise.org/projects/KIDS/}};][]{Jong_2013} -- and upcoming -- Nancy Grace Roman Space Telescope \citep[formerly known as the Wide Field Infrared Survey Telescope\footnote{\url{https://roman.gsfc.nasa.gov}};][]{Spergel_2015}, Euclid\footnote{\url{https://sci.esa.int/web/euclid}} \citep[]{Laureijs_2011}, and the Large Synoptic Survey Telescope \citep[LSST\footnote{\url{https://www.lsst.org}};][]{LSST_2009} -- surveys, which will observe significant fractions of the celestial sphere. For future Stage IV wide-field surveys, mass mapping must be constructed natively on the sphere or inverted allowing for the spherical forward model to form the Spherical Kaiser-Squires (SKS) estimator to avoid errors due to projection effects. In general, the Kaiser-Squires method is a widely known approach but it is not robust to noise, field boundary effects, or survey masks on the shear fields. \\

In this paper, the previously developed improved Kaiser-Squire \citep[KS+][]{Pires_2020} formalism is extended to the sphere (henceforth referred to as SKS+).
This paper is structured as follows. \S \ref{sec: Formalism}
describes the theoretical foundation of weak gravitational lensing, as well as data structures for spherical mapping, that is HEALPix and isotropic Undecimated Wavelet Transform on the Sphere (UWTS), following \cite{Bartelmann_2001}, \cite{ Castro_2005}, \cite{Gorski_2005} and \cite{Starck_2006}. In \S \ref{sec: Spherical Mass mapping reconstruction} the full methodology that was used to generate the weak lensing convergence maps from shear, that is the reconstruction of weak lensing convergence maps using SKS and SKS+, is described.
\S \ref{sec: Quantifying the quality of reconstruction} describes the quality of the SKS+ method using a two-point correlation function, power spectrum, the third- and fourth-order moments, and peak statistics of the reconstructed convergence maps along with the effects of masking, sampling, and noise, and a comparison of the spherical case to the planar setting is provided. Concluding remarks are made in \S \ref{sec: Summary and Conclusion}.

%%%%%%%%%%%%%%%%%%%%%%%%%%%%%%%%%%%%%%%%%%%%%%%%%%%%%%%%%%%%%%%%
%%%%%%%%%%%%%%%%%%%%%%%%%%%%%%%%%%%%%%%%%%%%%%%%%%%%%%%%%%%%%%%%
\section{Formalism}
\label{sec: Formalism}

%%%%%%%%%%%%%%%%%%%%%%%%%%%%%%%%%%%%%%%%%%%%%%%%%%%%%%%%%%%%%%%%
\subsection{\textbf{Weak lensing on Sphere}}
\label{sec: Weak lensing}
We start by laying out the formalism for obtaining the full-sky convergence field $\kappa$ as a set of spherical harmonics, constrained by observations of the shear $\gamma$ field \citep[e.g.][]{Bartelmann_2001, Castro_2005, Chang_2018}.
\begin{flushleft}
\justify
In gravitational weak lensing, the \textit{lensing potential}, $\phi$, at 3D position in comoving space \textbf{\textit{r}} $\equiv$ (\textit{r} , $\theta$, $\varphi$) is defined by a surface integral over the 2D projected gravitational potential, $\Phi$, along the line of sight \citep{Bartelmann_2001} as
\end{flushleft}

\begin{equation} \label{eq1}
  \hypertarget{equation 1} {}
    \phi (\textbf{\textit{r}}) =  \frac{2}{c^2} \int_{0}^{\textit{r}} dr' \frac{f_K (r-r')}{{f_K(r)} {f_K(r')}} \Phi (r' , \theta, \varphi),
\end{equation}

\justify
where coordinate $r$ is a radial distance, ($\theta, \varphi$) represents the angular position or spherical coordinates in the sky with colatitude $\theta \, \in \, [0, \pi]$ and longitude $\varphi \, \in \, [0, 2\pi]$ and $\mathit{f_{K} }$ is a comoving angular diameter distance which depends on the curvature \emph{K} of the universe.
\justify
The gravitational potential $\Phi$ is related to the matter overdensity, $\delta (\textbf{\textit{r}}) \equiv \delta \rho (\textbf{\textit{r}}) / \rho$ by the Poisson equation:

\begin{equation} \label{eq2}
  \hypertarget{equation 2} {}
   \begin{aligned}
    \nabla^{2}_r \Phi (\textbf{\textit{r}}) =  \frac{3 \Omega_m H_0^2}{2 a(t)} \delta (\textbf{\textit{r}}) 
  \end{aligned},
\end{equation}

\justify
where the 3D gradient $\nabla_r^{2}$ is defined relative to the comoving coordinates, $\Omega_m$ is the density of total matter to the present day, $H_0$ is the Hubble constant today in units of km/s/Mpc, and $a(t) = 1/(1 + z)$ is the scale factor.
\justify
The decomposition of the lensing potential \eqref{eq1} and its inverse into spherical harmonics is
\begin{equation} \label{eq3}
  \hypertarget{equation 3} {}
  \phi (\textbf{\textit{r}}) = \sum_{l=0}^{\infty} \sum_{m=-l}^{l}   \phi_{lm} Y_{lm} (\theta, \varphi),
\end{equation}

\begin{equation} \label{eq4}
  \hypertarget{equation 4} {}
   \phi_{lm} =\int d\Omega(\theta, \varphi) \phi (\textbf{\textit{r}}) Y_{lm}^{\star} (\theta, \varphi),
\end{equation}

\justify
where $d\Omega (\theta, \varphi) = \sin \theta \, d\theta \, d\varphi$ is the usual invariant measure on the sphere,
 $Y_{lm}$ are the spherical harmonics, $\phi_{lm}$ are the spherical harmonic coefficients and $\star$ denotes the complex conjugation, and $\mathit{l}$ and $\mathit{m}$ are the spherical harmonic degree and order, respectively.
\justify
Now, the spherical representation of the shear field can be defined as

\begin{equation} \label{eq5}
  \hypertarget{equation 5} {}
      \gamma(\textbf{\textit{r}}) = \gamma_{1} + i \gamma_{2} =\sum_{l=0}^{\infty} \sum_{m=-l}^{l}{ \hat{\gamma}_{lm}}  \prescript{}{\pm2}{Y_{lm} (\theta, \varphi)},
\end{equation}

\justify
where the real component, $\gamma_{1}$ , represents the elongation along the $\mathit{x}$ axis, and the compression along the $\mathit{y}$ axis, $\gamma_{2}$, represents a similar distortion, but rotated by $+45^{\circ}$.
\justify
The convergence in terms of harmonics expansion of the lensing potential $\phi$ on the sphere can be written as

\begin{equation} \label{eq6}
  \hypertarget{equation 6} {}
      \kappa(\textbf{\textit{r}}) = \kappa_{E} + i \kappa_{B} =\sum_{l=0}^{\infty} \sum_{m=-l}^{l} \hat{\kappa}_{lm} \prescript{}{0}{Y_{lm} (\theta, \varphi)},
\end{equation}
\justify
where $\prescript{}{s}{Y_{lm} (\theta, \varphi)}$ is the spin-weighted spherical harmonic and s is its spin weight. The $\hat{\gamma}_{lm}$ and $\hat{ \kappa}_{lm}$ are the spherical harmonic space coefficients associated with $\prescript{}{s}{Y_{lm} (\theta, \varphi)}$ at $\textbf{\textit{r}}$\footnote{we omitted the $\textbf{\textit{r}}$ reference in our notation for simplicity}.

\justify
The equations for shear and convergence \citep{Castro_2005, Chang_2018, Pichon_2010, Christopher_2017} are:
\begin{equation} \label{eq7}
  \hypertarget{equation 7} {}
  {\hat{\gamma}_{lm}}\prescript{}{\pm2} \,= \, \hat{ \gamma}_{lm} = \hat{\gamma}_{E,lm} + i \hat{\gamma}_{B,lm} = \frac{1}{2} [l (l+1) (l-1) (l+2)]^{ \frac{1}{2}} \phi_{lm}
\end{equation}

\begin{equation} \label{eq8}
  \hypertarget{equation 8} {}
  {\hat{\kappa}_{lm}}\prescript{}{0} \,=\, \hat{\kappa}_{lm} = \hat{\kappa}_{E,lm} + i \hat{\kappa}_{B,lm} = - \frac{l(l+1)}{2} \phi_{lm}.
\end{equation}

\justify
From Eq. \eqref{eq7} and Eq. \eqref{eq8}, the derived relation between shear $\gamma$ and convergence $\kappa$ is represented as

\begin{equation} \label{eq9}
  \hypertarget{equation 9} {}
   \hat{\gamma}_{lm}= - \sqrt{\frac{(l+2)(l-1)}{l(l+1)}} ( \hat{\kappa}_{E,lm} + i \hat{\kappa}_{B,lm}) = S_l \, \hat{\kappa}_{lm}
\end{equation}
which is the spherical forward model, where $S_l = - \sqrt{\frac{(l+2)(l-1)}{l(l+1)}}$.
\justify
The E and B modes of the convergence field can be calculated by applying Eq. \eqref{eq9} on the spherical harmonics of shear. 
\justify 
In the flat sky limit, the decomposition into spherical harmonics is replaced by the Fourier transform,
 \begin{equation}
    \sum \phi_{\mathit{lm}} Y_{lm}  \to \int \frac{d^2 l}{(2 \pi)^2} \phi(\mathit{l}) \mathit{e}^{il\cdot\theta}
 \end{equation}

\justify
The above equations then reduce to the usual \citet{KS_1993} formalism, and this mapping can be referred to as Spherical Kaiser-Squires (SKS).

%%%%%%%%%%%%%%%%%%%%%%%%%%%%%%%%%%%%%%%%%%%%%%%%%%%%%%%%%%%%%%%%
\subsection{\textbf{HEALPix}}
\label{sec: Healpix}

Hierarchical Equal Area isoLatitude Pixelisation \citep[HEALPix\footnote{\url{https://healpix.jpl.nasa.gov}};][]{Gorski_2005} is a scheme to pixelize the curved sky. The aim of HEALPix is to support fast numerical analysis of data over the sphere. The notion is to support convolutions with local and global kernels, Fourier analysis with spherical harmonics, power spectrum estimation and topological analysis. The primary benefit of the HEALPix scheme is that all pixels have equal areas because the white noise generated by the signal receiver gets integrated exactly into the white noise in the pixel space. However, this comes at the cost of each pixel having a different shape.
\justify
The variable $N_{side}$ parameter describes the resolution of HEALPix maps and also determines the number of pixels in the sphere as $N_{pixel} = (12N_{side}^2)$ of the same area $A_{pix} = \frac{\pi}{3N_{side}^2}$. Throughout this article, the maps are constructed using ${N_{side}} = 2048$, corresponding to a pixel size of 1.7177 arcmin and all relevant spherical harmonic transforms use $l_{max} = (3 {N_{side}} - 1)$.
\justify
Spherical Harmonic Transforms\footnote{The Fourier transform, in the flat space, is the 'harmonic transform'. It is the Laplacian's expansion on the curved space in terms of eigen functions.} are non-commutative Fourier transforms on the sphere.
The HEALPix function $map2alm$ is used to decompose the shear field into a spherical harmonic space obtaining the coefficients $\hat{\gamma}_{E, lm}$, $\hat{\gamma}_{B, lm}$ and calculate $\hat{\kappa}_{E, lm}$, $\hat{\kappa}_{B, lm}$ according to Eq. \eqref{eq9}.
Finally, the HEALPix function $alm2map$ is used to convert these coefficients back to the real-space ${\kappa}_{E}$ and ${\kappa}_{B}$ maps.
The HEALPix routine $anafast$ is used to estimate the power spectra of the maps.

%%%%%%%%%%%%%%%%%%%%%%%%%%%%%%%%%%%%%%%%%%%%%%%%%%%%%%%%%%%%%%%%
\subsection{\textbf{Isotropic Undecimated Wavelet Transform on the Sphere}}
\label{sec:UWTS}

An Isotropic Undecimated Wavelet Transform on the Sphere (UWTS) based on spherical harmonics was described by \citet{Starck_2006} that is used to reconstruct the wide field convergence maps.
This isotropic transform is based on  scaling function $\phi_{l_c} (\theta, \varphi)$ with cut-off frequency $\mathit{l_{c}}$ and which depends only on co-latitude $\theta$ and is invariant with respect to a change in longitude $\varphi$. The spherical harmonic coefficients $\phi_{l_c} (\mathit{l}, \mathit{m})$ of $\phi_{l_c}$ vanish when $\mathit{m} \, \neq \, 0$ which makes it simple to compute the spherical harmonic coefficients $\hat{c_0} (\mathit{l}, \mathit{m})$ of $\mathit{c_0} = \phi_{l_c} \, \ast \, \mathit{f}$, where $\ast$ stands for convolution.

\begin{equation} \label{eq10}
  \hypertarget{equation 10} {}
     \hat{c_0}(l,m) = \widehat { \phi_{l_c} * f} (l,m) = \sqrt{\frac{4 \pi}{2l+1}} \hat{\phi}_{l_c} (l,0) \hat{f}(l,m) \\
\end{equation}
\justify
Using this scaling function ($\phi_{l_c}$), it is possible to get a sequence of smoother approximations $\mathit{c_{j}}$ of the function $\mathit{f}$ on a dyadic resolution scale. Let $\phi_{2^{-j} l_c}$ be a rescaled version of $\phi_{l_c}$ with cut-off frequency $2^{-j}\mathit{l_{c}}$. Then the multiresolution decomposition $\mathit{c_{j}}$ can be obtained by convolving $\mathit{f}$ with $\phi_{2^{-j} l_c}$ \\
\begin{equation} \label{eq11}
  \hypertarget{equation 11} {}
  \begin{aligned}
     c_0 &= \phi_{l_c} * f \\
     c_1 &= \phi_{2^{-1} l_c} * f = c_0 * h_0 \\
     &. . . \\
     c_j &= \phi_{2^{-j} l_c} * f = c_{j-1} * h_{j-1} \\
  \end{aligned}
\end{equation}
\justify
where the zonal low pass filters $\mathit{h_{j}}$ are defined by

\begin{equation} \label{eq12}
  \hypertarget{equation 12} {}
    \begin{aligned}
     \hat{H_j}(l,m) &= \sqrt{\frac{4 \pi}{2l+1}} \hat{h_j} (l,m) \\ &= \begin{cases}
    \frac{\hat{\phi}_{\frac{l_c}{2^{(j+1)}}}(l,m)}{\hat{\phi}_{\frac{l_c}{2^j}}(l,m)},& \text{if } l<\frac{l_c}{2^{(j+1)}}\, \text{and}\, m = 0 \\
    0,              & \text{otherwise}
   \end{cases}
  \end{aligned}
\end{equation}
\justify
The possible scaling function\footnote{This choice is motivated by the desire to analyse a function that is close to a Gaussian while also validating the dilation equation, which is required for a fast transformation.} is $\phi_{l_c} (\mathit{l}, \mathit{m}=0) = \frac{2}{3}$B\textsubscript{3}($\frac{2l}{l_{c}}$), where B\textsubscript{3} is the cubic B-spline

\begin{equation} \label{eq13}
  \hypertarget{equation 13} {}
    \begin{split}
     B_{3}(x)= \frac{1}{12}(\mid x-2 \mid^3 - 4\mid x-1 \mid^3 + 6\mid x \mid^3 - \\
     4\mid x+1 \mid^3 + \mid x+2 \mid^3) 
    \end{split}
\end{equation}
\justify
Just like with the \`a trous algorithm by \citet{Starck_2010}, the wavelet coefficients $\mathit{w_{j}}$ can be defined as the difference between two consecutive resolutions.

\begin{equation} \label{eq14}
  \hypertarget{equation 14} {}
     w_{j+1}(\theta, \varphi)= c_j(\theta, \varphi) - c_{j+1}(\theta, \varphi) \\
\end{equation}
\justify
which are equivalent to the wavelet function $\psi_{\mathit{l_{c}}}$

\begin{equation} \label{eq15}
  \hypertarget{equation 15} {}
    \hat{\psi}_{\frac{l_c}{2^j}}(l,m) = \hat{\phi}_{\frac{l_c}{2^{j-1}}}(l,m) - \hat{\phi}_{\frac{l_c}{2^j}}(l,m) \\
\end{equation}
\justify
so that
\begin{equation} \label{eq16}
  \hypertarget{equation 16} {}
  \begin{aligned}
     w_0 &= \psi_{l_c} * f \\
     w_1 &= \psi_{2^{-1} l_c} * f  \\
     &. . . \\
     w_j &= \psi_{2^{-j} l_c} * f  \\
  \end{aligned}
\end{equation}
\justify
Since the wavelet coefficients are defined as the difference between two resolutions, the reconstruction from the wavelet decomposition $\mathit{W} = \{\mathit{w_{1}}, . . . , \mathit{w_{j}}, \mathit{c_{j}}$ \} is straightforward and corresponds to the reconstruction as of the \`a trous algorithm
\begin{equation} \label{eq17}
  \hypertarget{equation 17} {}
     f(\theta, \varphi)= c_j(\theta, \varphi) + \sum_{j=1}^{J} w_j(\theta, \varphi) \\
\end{equation}
\justify
where the simplifying assumption has been made that $\mathit{f}$ is equal to $\mathit{c_{0}}$ and $\mathit{J}$ represents the number of wavelet bands/scale.

%%%%%%%%%%%%%%%%%%%%%%%%%%%%%%%%%%%%%%%%%%%%%%%%%%%%%%%%%%%%%%%%%%%%%%%%%%%%%%%%%%%%%%%%%%%%%%%%%%%%%%%%%%%%%%%%%%%%%%%%%%%%%%%%
\section{Spherical mass inversion}
\label{sec: Spherical Mass mapping reconstruction}

In this section, the process of estimating the convergence map from an observed shear field on the sphere using SKS and SKS+ methods is discussed.

%%%%%%%%%%%%%%%%%%%%%%%%%%%%%%%%%%%%%%%%%%%%%%%%%%%%%%%%%%%%%%%%
\subsection{\textbf{Kaiser-Squires on the sphere (SKS)}}
\label{sec: Kaiser-Squires on the sphere}
\justify
The reconstruction of the convergence map on the sphere from weak lensing involves solving a spherical inverse problem, as discussed above.
In the weak lensing regime, the observational constraint can be fully expressed in terms of the reduced shear, $g$.
\begin{equation}
    g \equiv \frac{\gamma}{1-\kappa}
\end{equation}

However, when mapping the large-scale dark matter distribution within the weak regime, the reduced shear is approximately the true shear, $\gamma \simeq g$. Therefore, throughout the article, it has been avoided to perform reduced shear iterations.

This allows the observed shear to be

\begin{equation} \label{eq19}
  \hypertarget{mask_gamma_noise} {}
    \gamma^{obs}  = M \gamma^n
\end{equation} 

\justify
where $\gamma^n = \gamma + \epsilon_s$ and '$\epsilon_s$' represent the shape noise described in \S \ref{sec:Shape noise}, subscript 'obs' refers to the observations, $\gamma$ the shear maps, and $\mathit{M}$ is the mask (explained in \S \ref{sec: Missing data} that correspond $\mathit{M}$ = 1 when measurements($\gamma^{n}$) have valid data and $\mathit{M}$ = 0 indicating invalid data in observations.
Therefore, Equation \ref{eq19} can be interpreted as a noisy measurement of true shear degraded by shape noise (caused by the unknown intrinsic ellipticities '$\epsilon_s$' of the observed galaxies) and Mask. \\

\justify
The Kaiser-Squires on the sphere uses a spin transformation of the shear maps $\gamma$ ($\gamma_E$ and $\gamma_B$) to get both the curl-free E-mode convergence map $\kappa_E$, which carries most of the cosmological signal, and the divergence-free B-mode convergence map $\kappa_B$, which arises due to non-linear lensing corrections. The HEALPix functions are $map2alm$ performs the decomposition of real space ($\gamma_E$ and $\gamma_B$) into spherical harmonic spaces ($\hat{\gamma}_{\mathit{E}, \mathit{lm}}^{obs}$  and $\hat{\gamma}_{\mathit{B},\mathit{lm}}^{obs}$) as in Eq. \eqref{eq9}, which can be re-written as

\begin{equation} \label{sks}
    %\hat{\kappa}_{lm}^{SKS} = M S_l^{-1} \hat{\gamma}_{lm}^{n} = S_l^{-1} \hat{\gamma}_{lm}^{obs}
    \hat{\kappa}_{lm}^{n} = M S_l^{-1} \hat{\gamma}_{lm}^{n} = S_l^{-1} \hat{\gamma}_{lm}^{obs}
\end{equation} 

\justify
then $alm2map$ converts the spherical harmonic space coefficients ($\hat{\kappa}_{\mathit{E}, \mathit{lm}}^{n}$ and $\hat{\kappa}_{\mathit{B}, \mathit{lm}}^{n}$) back to the real space maps ($\kappa_E$ and $\kappa_B$).
However, because this method does not account for noise and survey masks on the shear fields, poor results are inevitable.

%%%%%%%%%%%%%%%%%%%%%%%%%%%%%%%%%%%%%%%%%%%%%%%%%%%%%%%%%%%%%%%%
\subsection{\textbf{Improved Kaiser-Squires on the sphere (SKS+)}}
\label{sec: Improved Kaiser-Squires}
\justify
In practice, due to the limitations of the observational quality i.e. noise (\S \ref{sec:Shape noise}) and survey mask windows, where noise covariance is typically large in magnitude, the inverse problem become strongly ill-posed and thus significant regularisation is required to stabilise the inversion.
The methods described in \citet{Pires_2020} have been extended to the sphere in this paper to incorporate the necessary corrections for imperfect and realistic measurements.
The convergence $\boldsymbol{\kappa}$ can be recovered by using a transformation $\boldsymbol{\Phi}$ which yields $\boldsymbol{\kappa}$ = $\boldsymbol{\Phi}\boldsymbol{\alpha}$. This deceptively simple expression requires some unpacking. In the spherical domain, the chosen dictionary $\boldsymbol{\Phi}$ is denoted as the spherical harmonic basis, conversely $\boldsymbol{\Phi^T}$ is the spherical harmonic transform, i.e. the projector onto the spherical harmonic space. The vector $\boldsymbol{\alpha}$=\{$\phi_{\mathit{lm}}$\}\textsubscript{$\mathit{l=0,1..l_{max}, m=-l,..,l}$} is the set of spherical harmonic coefficients that reproduce the convergence, $\kappa$.\\

The extension of KS+ \citep{Pires_2020} can accommodate missing data to the celestial sphere. After applying SKS+, the missing data regions in the reconstructed convergence map and the corresponding shear maps have evolved to non-zero values. 
Then, following the essential idea of \citet{Pires_2020}, the best solution is identified from the set of all possible solutions as the sparsest, that is, the solution with the spherical harmonic coefficients, $\boldsymbol{\alpha}$.\\

\justify
The mass-inversion can be estimated by solving the following.
\begin{equation} \label{MInversion}
  \hypertarget{equation 20} {}
  % \min_{\alpha} \parallel \alpha \parallel_{0} \text {subject to}\, \parallel \gamma^{obs} -  M \gamma^n  \parallel^2 \;\; \leq \sigma
  \min_{\boldsymbol{\hat{\kappa}_{lm}^{n}}} \parallel \boldsymbol{\Phi^T} \, \boldsymbol{\hat{\kappa}_{lm}^{n}} \parallel_{0} \text {subject to}\, \parallel \boldsymbol{\hat{\gamma}_{lm}^{obs} -  M \, S_l \, \hat{\kappa}_{lm}^{n}}  {\parallel}_{l_2} \; \leq \sigma  
\end{equation}
where ${\parallel \cdot \parallel}_0$ is the pseudonorm, that is, the number of non-zero entries in ($\cdot$),  $\parallel \cdot \parallel$ the classical $l_2$ norm (i.e. $\parallel \cdot \parallel = \sqrt{\sum_k (\cdot_k)^2})$ and $\sigma$ stands for the noise standard deviation of the input shear map measured outside the mask.
The solution to this convex optimisation problem can be obtained using an iterative thresholding process \citep{Starck_2004} and modifying it by multiplying the full residual by $\mathit{M}$ after each residual estimation as in \citet{Abrial_2008}.
\justify

In ideal lens-source clustering \citep[see][]{Schneider_2002_2PCF}, all astronomical weak lensing only approximately gives pure E modes and, in practice, observational effects, due either to an imperfect point-spread function correction or spatially varying calibration errors under the assumption that the PSF correction is correct or to an intrinsic alignment of galaxies arising from spin-spin correlation, induce B modes.
The consequence is that any observational effects that would normally go into B modes are necessarily pumped into E modes. Thus the B-modes can provide an important check for systematic errors. In this paper, we force the B modes to zero inside the mask rather than letting them free, and we incorporate an additional constraint on the power spectrum of the convergence map as of \citet{Pires_2020}.
The way this is implemented is using the Isotropic Undecimated Wavelet Transform on the Sphere.
As described in \hyperref[sec:UWTS]{section 2.3}, the idea is to decompose our image $\kappa$ into a set of coefficients \{$\mathit{w_{1}}, \mathit{w_{2}}, . . . , \mathit{w_{j}}, \mathit{c_{j}}$\}, as a superposition of the form Eq. \eqref{eq17} (in this equation, $\mathit{f}$ is equivalent to $\kappa$) where $\mathit{c_{j}}$ is a smoothed version of the image, and $\mathit{w_{j}}$ are a set of aperture mass maps (usually called wavelet bands) at scale $\theta = \mathit{2}^{j}$. Aperture mass maps, $\mathit{w_{j}}$, are obtained by convolving the image $\kappa$ with 1D filter $h_j$ derived from the B3-spline as shown in Eq. \eqref{eq12} and \eqref{eq13}.
Then, the variance of each wavelet band $\mathit{w_{j}}$ is estimated. The variance per scale estimated in this way can be directly compared to the power spectrum. \\
\justify
This provides a way to estimate a broadband power spectrum of the convergence $\kappa$ from incomplete data. The power spectrum is then enforced by multiplying each wavelet coefficient by the factor ${\sigma}_{{j}}^{out}$/${\sigma}_{{j}}^{in}$ inside the gaps and then reconstructing the convergence through the inverse wavelet transform. The ${\sigma}_{{j}}^{in}$ and ${\sigma}_{{j}}^{out}$ are the standard deviation estimated in the wavelet band $\mathit{w_{j}}$ inside and outside the mask, respectively.\\
\justify
After implying constraints, the equation (\ref{MInversion}) that we want to minimise can be re-written as \\
  \begin{multline} \label{KSP}
  \min_{\boldsymbol{\hat{\kappa}_{lm}^{n}}} \parallel \boldsymbol{\Phi^T} \, \boldsymbol{\hat{\kappa}_{lm}^{n}} \parallel_{0} \;
   \text {subject to} \\ \parallel \boldsymbol{\hat{\gamma}_{lm}^{obs} -  M \, S_l \, W^T \, P \, W \, \hat{\kappa}_{lm}^{n}}  {\parallel}_{l_2} \; \leq \sigma 
  \end{multline}

\justify
where P is a linear operator to impose power spectrum constraint, $W$ and $W^T$ are the forward wavelet transform and inverse wavelet transform on the sphere respectively (see algorithm \ref{euclid}\footnote{The algorithm depends (only) on the choice of the $N_{side}$ parameter, which sets the number of pixels, pixel area and the spherical harmonic degree and order.}).\\
\justify
We quantified the quality of the reconstructed convergence maps with respect to missing data and shape noise in \S \ref{sec: Quantifying the quality of reconstruction}.

\renewcommand{\labelenumii}{\roman{enumii}}
        \begin{breakablealgorithm} \footnotesize
         \caption{SKS+ algorithm}\label{euclid}
        \begin{enumerate}
            \item \textbf{Input:} Shear map $\gamma$ with mask in HEALPix format.
        \item Compute ${\kappa}$ by performing a direct spherical mass inversion. \S \ref{sec: Kaiser-Squires on the sphere}
       \item Set the number of iterations $I_{\rm{max}}$, the maximum threshold $\lambda_{\rm{max}}= \max(|\alpha=\Phi^T \kappa_E|)$, and the minimum threshold $\lambda_{min} = 0$. 
        \item Initialise solution: ${\kappa}^0$ = 0
        \item Set $\mathit{i}$ to 0, $\lambda_{i}$ = $\lambda_{max}$ and ${\kappa}^i$=${\kappa}$. Iterate:
        \begin{enumerate}
          		\item Compute $\alpha$=$\Phi^T$${\kappa}^i$
          		\item Compute $\tilde{\alpha}$ by setting to zero the coefficients $\alpha$ below the threshold $\lambda_{i}$.
                          \item  Reconstruct ${\kappa}^i$ from $\tilde{\alpha}$: ${\kappa}^i$= $\Phi$$\tilde{\alpha}$.
                           \item  Decompose ${\kappa_E}^{i}$ into its wavelet coefficients \{$\mathit{w}_{1}, \mathit{w}_{2}, ......, \mathit{w}_{j}, \mathit{c}_{j}$\}.
                           \item  Renormalise the spherical wavelet coefficients $\mathit{w}_{j}$, by a factor ${\sigma}_{{j}}^{out}$/${\sigma}_{{j}}^{in}$ inside the gaps.
                          \item   Reconstruct ${\kappa}^i$ by performing the backward spherical wavelet transform from the normalised coefficients.
                          \item   Perform the inverse mass relation to compute ${\gamma}^{i}$
                           \item  Enforce the observed shear ${\gamma}$ outside the gaps: \\ ${\gamma}^{i}$ = $\mathit{(1-M)}$ ${\gamma}^{i}$ + $\mathit{M}$  $\gamma$.
                          \item  Perform the direct mass inversion and get ${\kappa}^i$
                           \item   Set $\mathit{i}$=$\mathit{i+1}$. 
                           \item   Update the threshold: ${\lambda}_i$ = $\mathit{F}$ ($\mathit{i}$, $\lambda_{min}$, $\lambda_{max}$) 
                            \item If $\mathit{i}$ < $\mathit{I}_{max}$, return to step i.
        \end{enumerate}
         \end{enumerate}
      \end{breakablealgorithm}

The function $\mathit{F}$ fixes the decrease in the threshold, which decreases exponentially at each iteration from a maximum value to zero as described in \cite{Pires_2009}.\\
\footnotesize
$\mathit{F} ( \mathit{i}, \lambda_{\rm{min}}, \lambda_{\rm{max}}) = \lambda_{\rm{min}} + (\lambda_{\rm{max}} - \lambda_{\rm{min}} ) (\mathit{1} - \mathit{\rm{erf}} (\mathit{2.8 \, i} / \mathit{l_{max}}$)) \\

 \begin{figure}[ht]
 \hypertarget{NbIter} {}
 \subfigure{\includegraphics[width=90mm]{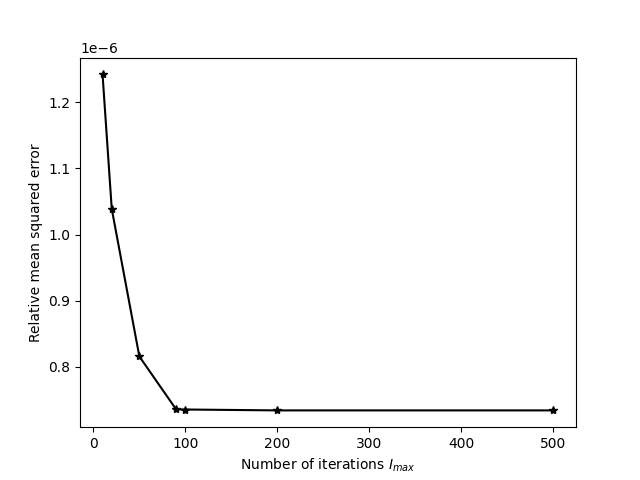}}
 \caption{Relative MSE as a function of maximum number of iterations $I_{max}$.}
 \label{Fig: NbIter}
 \end{figure}

We applied this algorithm with around $I_{max} = 100$ iterations in the spherical harmonics representation. We estimated the relative mean squared error (MSE) for various $I_{max}$ values to set this parameter. Figure \ref{Fig: NbIter} depicts the relative MSE per $I_{max}$ and we discovered that the error does not decrease significantly after $I_{max} > 90$, so we used $I_{max} = 100$.
We performed the test with other resolution maps and reached similar conclusions.

%%%%%%%%%%%%%%%%%%%%%%%%%%%%%%%%%%%%%%%%%%%%%%%%%%%%%%%%%%%%%%%%
%%%%%%%%%%%%%%%%%%%%%%%%%%%%%%%%%%%%%%%%%%%%%%%%%%%%%%%%%%%%%%%%
\section{Data}
\label{sec: Data}

Convergence map reconstruction using the SKS+ method has been tested on several shear maps generated by high resolution and realistic N-body simulations \citep{Takashi_2017}\footnote{The dataset used in this article can be found at \url{http://cosmo.phys.hirosaki-u.ac.jp/takahasi/allsky_raytracing}}. These simulation data sets are generated by performing multiple lens plane ray-tracing through high-resolution cosmological N-body simulations for redshifts z $\epsilon$ [0.05, 5.3] at intervals of 150 $h^{-1}$ Mpc comoving radial distance. Each box has $2048^3$ Dark Matter particles, so the resolution is higher in boxes closer to the observer.
The cosmological parameters selected for this suite of simulations are matter density $\Omega_m$ = 1, cosmological constant $\Omega_\Lambda$ = 0.279, baryon density $\Omega_b$ = 0.046, Hubble parameter $h$ = 0.7, normalisation  $\sigma_8$ = 0.82 and slope of the primordial power spectrum $n_s$ = 0.97. These parameters were adopted from standard cosmology $\Lambda$CDM (Lambda cold dark matter) that is consistent with the WMAP 9-year result \citep{Hinshaw_2013}.\\

\begin{figure*}
 \hypertarget{Figure 2} {}
\subfigure{\includegraphics[width=90mm]{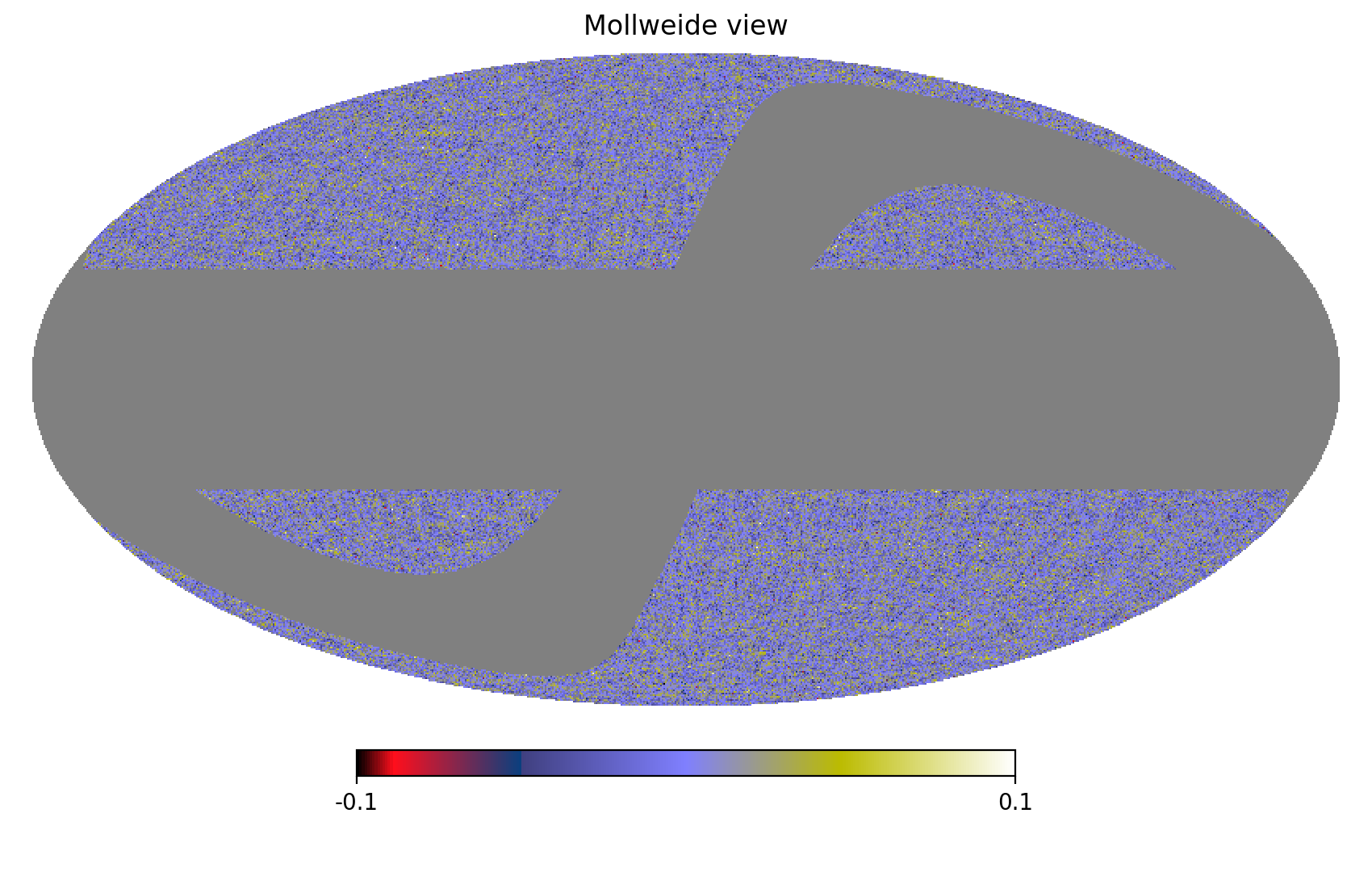}}
\subfigure{\includegraphics[width=90mm]{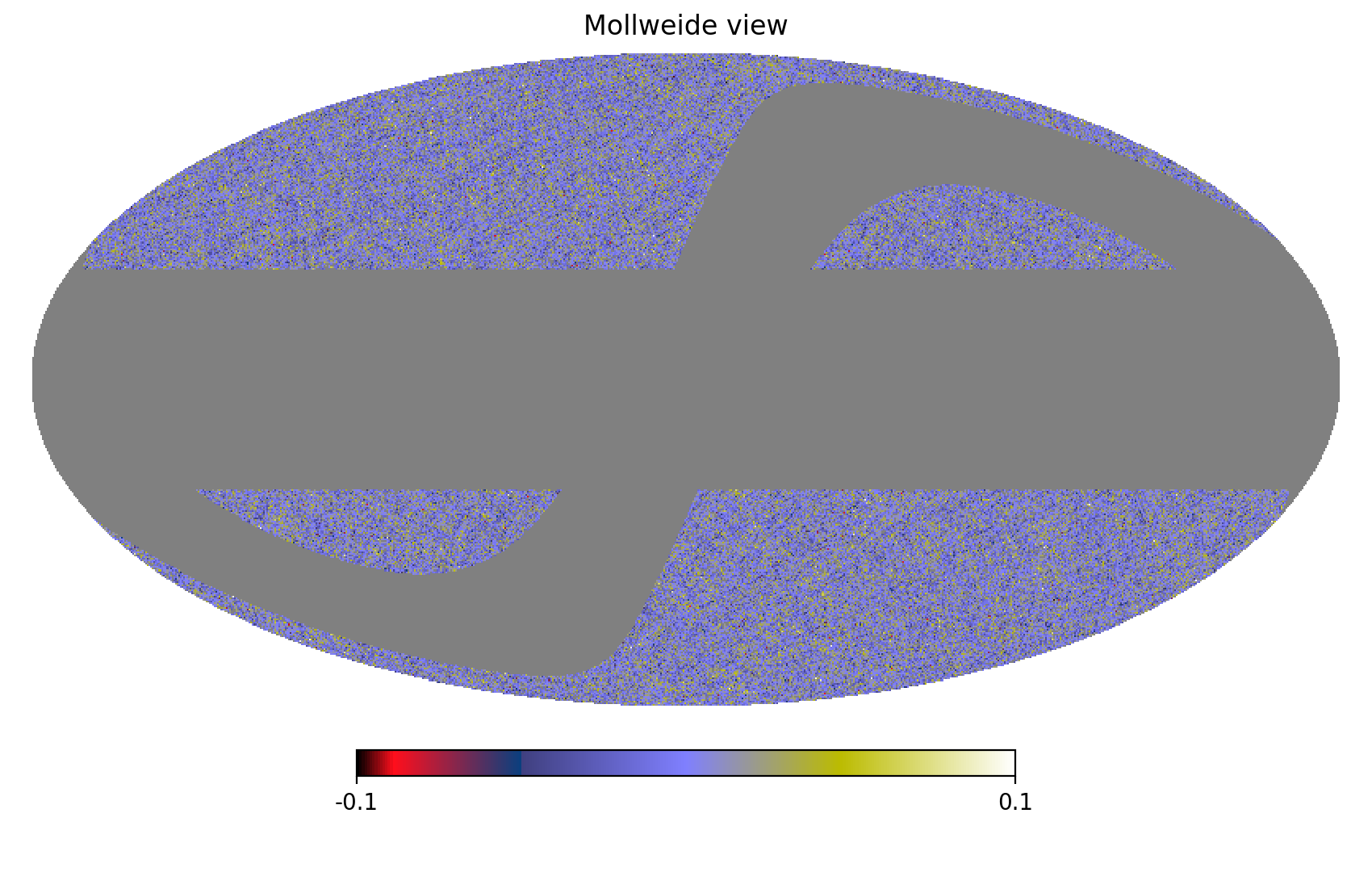}}
 \subfigure{\includegraphics[width=90mm]{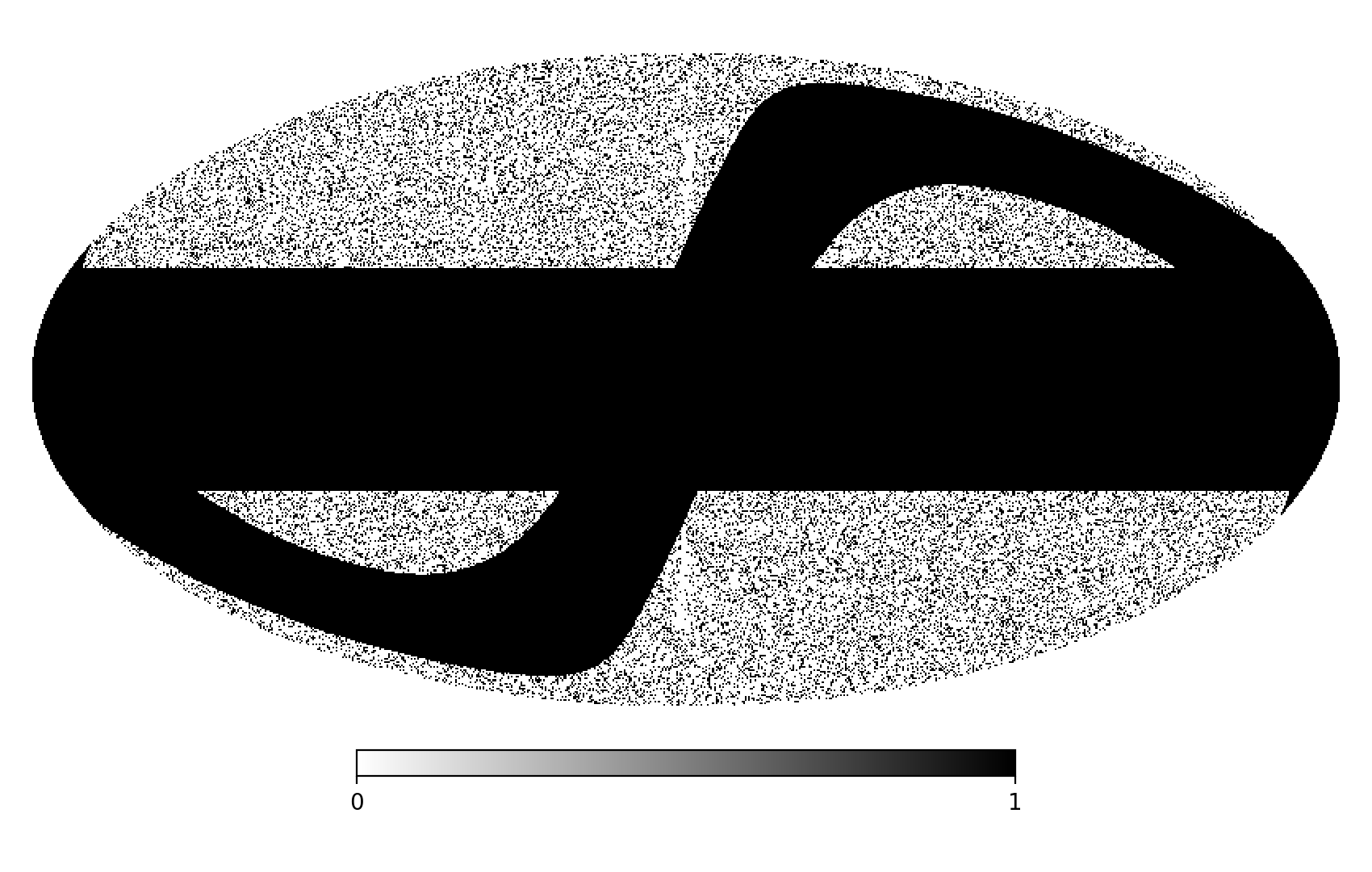}}
 \subfigure{\includegraphics[width=90mm]{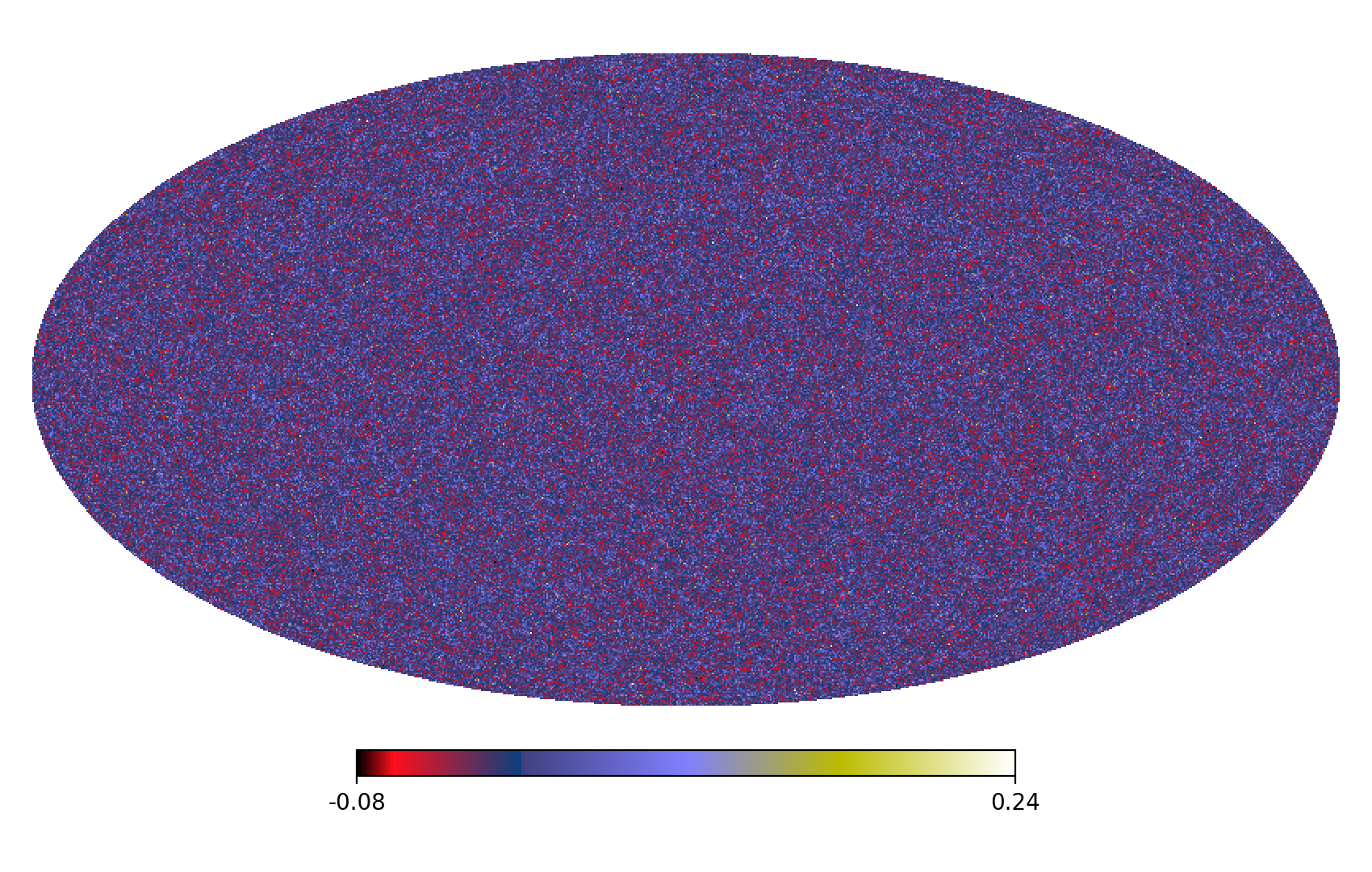}}
 \subfigure{\includegraphics[width=90mm]{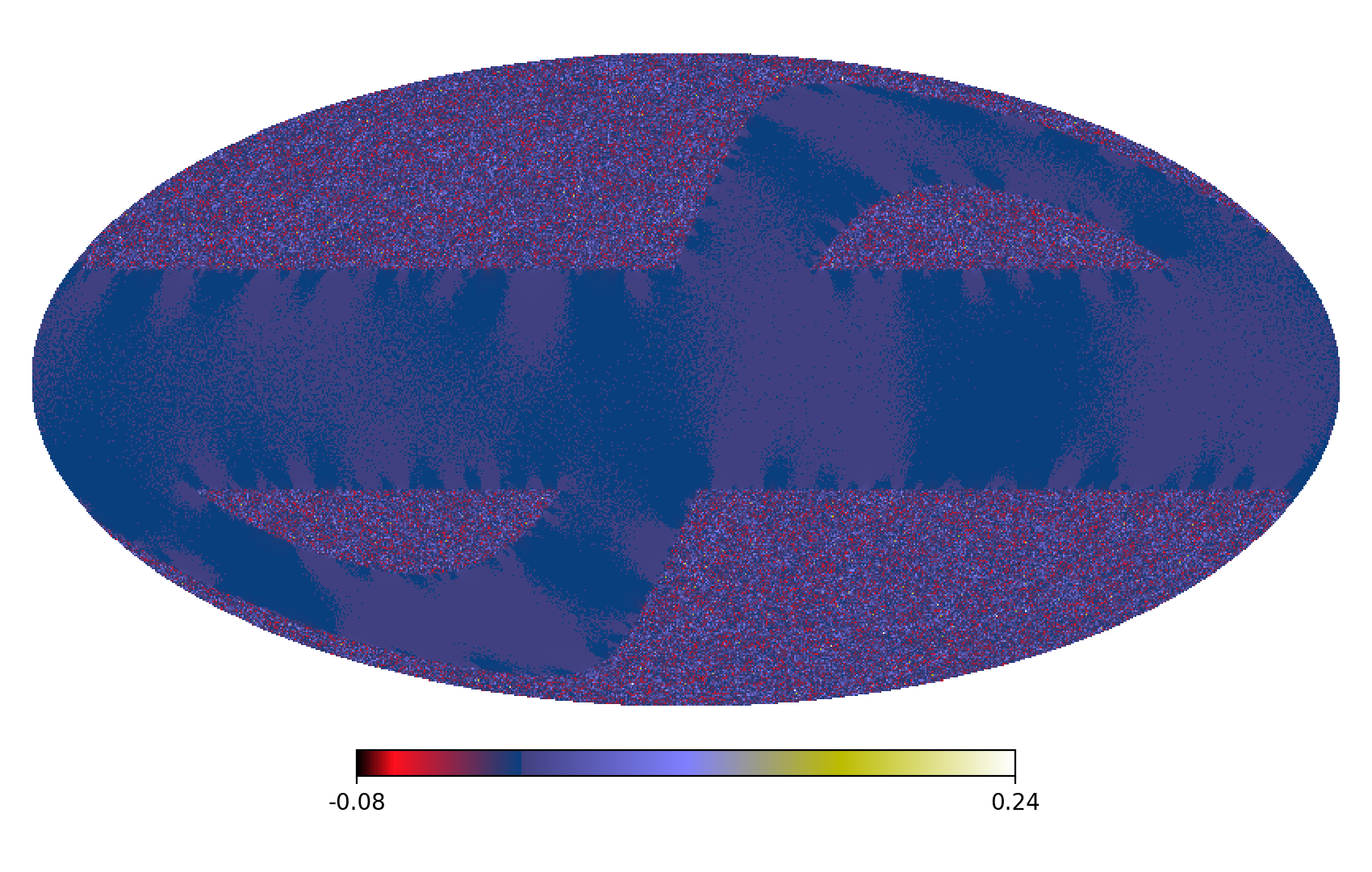}}
 \subfigure{\includegraphics[width=90mm]{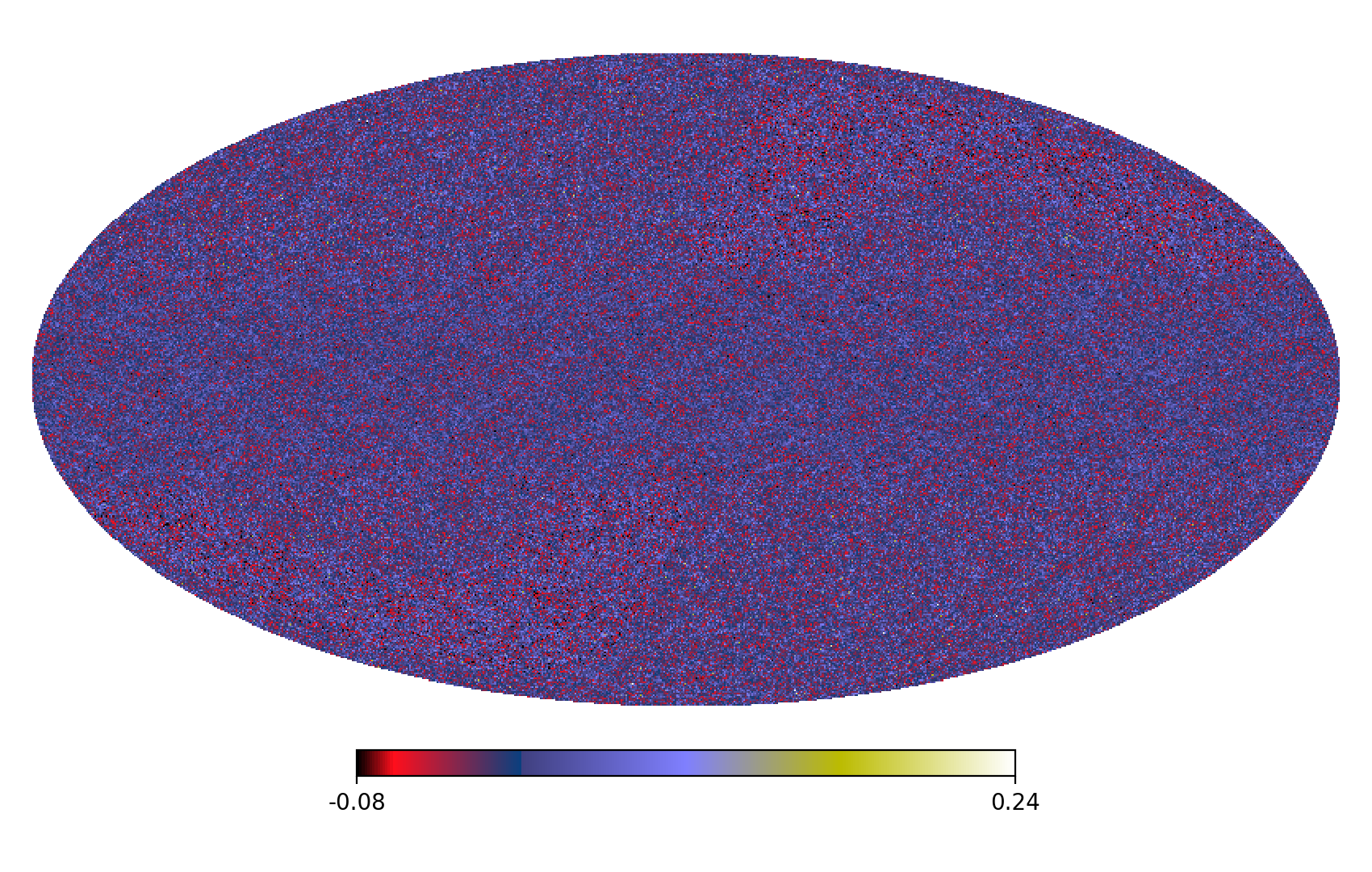}}

\caption{Shear field is shown in the first and second columns of the upper panel (the first showing ${\gamma_1}$ and the second showing ${\gamma_2}$) with missing data. The first column of the middle panel shows the mask which applied to the shear maps and the second column shows the original E-mode convergence ${\kappa}$ map.
The Lower panel shows the E-mode convergence maps reconstructed from the incomplete shear field using the SKS method (left), and using the SKS+ method (right).}
  \label{shear-convergence}
\end{figure*}

For the purpose of this article, shear maps at $z=2.9367$ which is close to the current limits of the stage IV survey. Shear maps are available in HEALPix pixelisation with a resolution of $N_{side}=4096$, which is degraded to a lower resolution map of $N_{side}=2048$ by using the HEALPix ud$\_$grade function (degrade by averaging the four input map pixels nested in one output map pixel value). Higher-resolution maps may produce better results at the cost of computational efficiency. \\
\justify
Finally, a pseudo-Euclid mask (masking of the stellar, galactic plane and the ecliptic), so as to best match upcoming Stage IV surveys like Euclid, has been applied to the shear maps as shown in \ref{shear-convergence}.

%%%%%%%%%%%%%%%%%%%%%%%%%%%%%%%%%%%%%%%%%%%%%%%%%%%%%%%%%%%%%%%%
\section{Quantifying the quality of reconstruction}
\label{sec: Quantifying the quality of reconstruction}
This section discusses the impact of missing data, shape noise, and projection on the SKS and SKS+ inversion methods and provides quantitative evidence of the efficacy of the SKS+ method.
%%%%%%%%%%%%%%%%%%%%%%%%%%%%%%%%%%%%%%%%%%%%%%%%%%%%%%%%%%%%%%%%

%%%%%%%%%%%%%%%%%%%%%%%%%%%%%%%%%%%%%%%%%%%%%%%%%%%%%%%%%%%%%%%%

%%%%%%%%%%%%%%%%%%%%%%%%%%%%%%%%%%%%%%%%%%%%%%%%%%%%%%%%%%%%%%%%

%%%%%%%%%%%%%%%%%%%%%%%%%%%%%%%%%%%%%%%%%%%%%%%%%%%%%%%%%%%%%%%%
\begin{figure*}[ht]
 \hypertarget{Residual maps} {}
 \subfigure{\includegraphics[width=90mm]{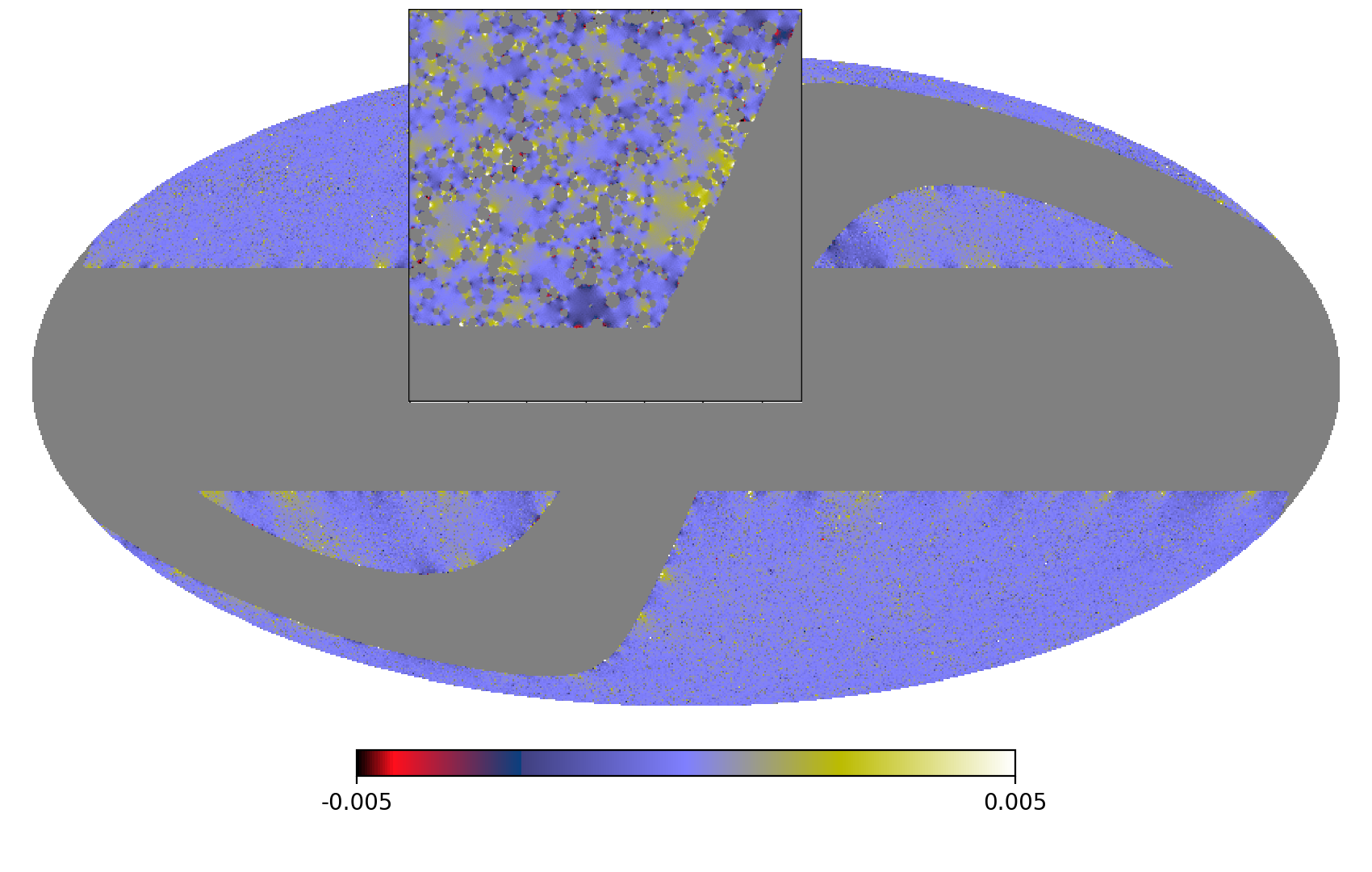}}
  \subfigure{\includegraphics[width=90mm]{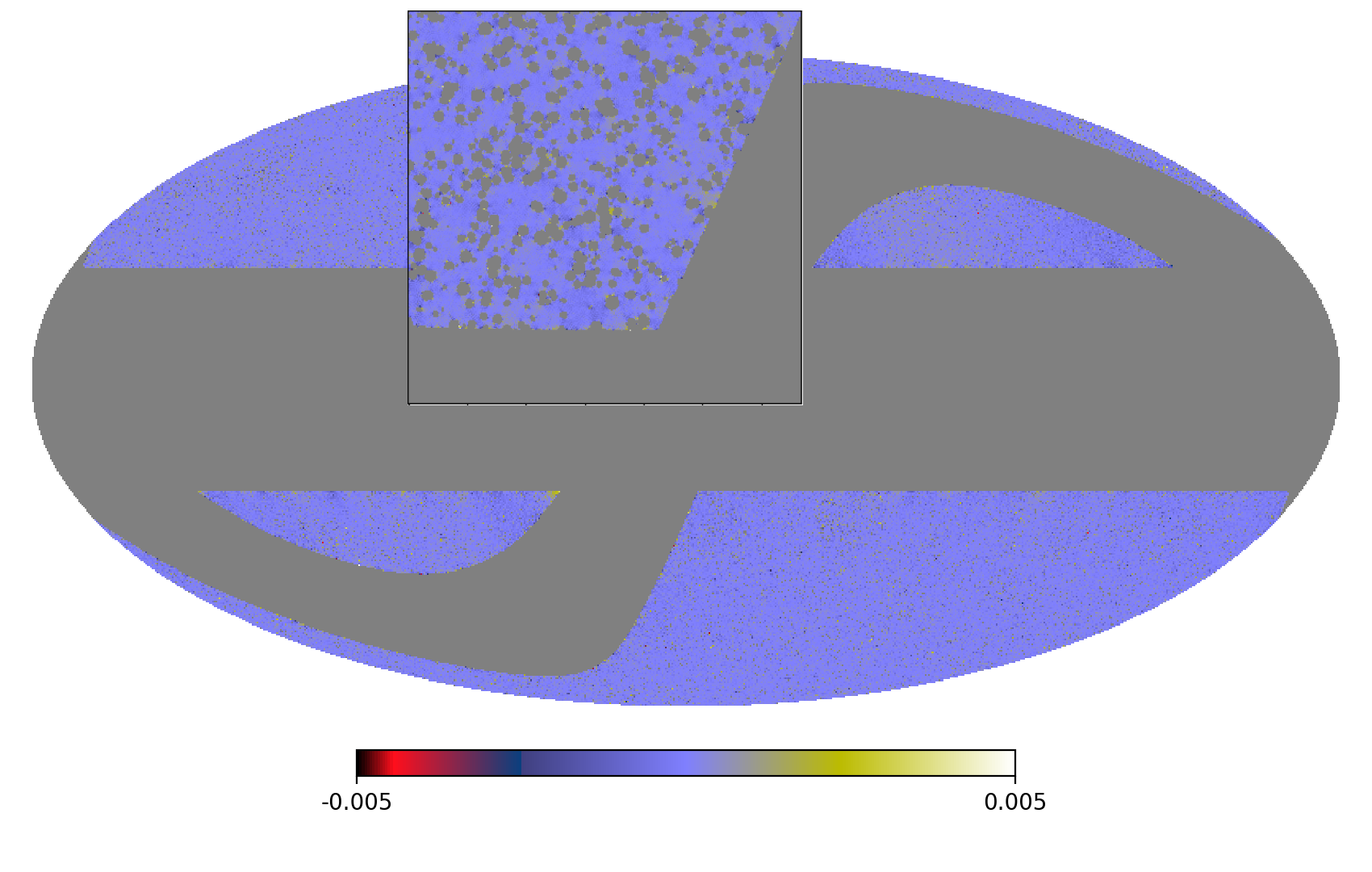}}
 \caption{Missing data effects: Pixel difference outside the mask between the original E-mode convergence map and the reconstructed E-mode convergence map using the SKS method (left) and the SKS+ method (right); the inner box corresponds to a zoom of the field $10^{\circ} \times 10^{\circ}$ with centre at (5, 28).}
   \label{fig:Residual maps}
 \end{figure*}

%%%%%%%%%%%%%%%%%%%%%%%%%%%%%%%%%%%%%%%%%%%%%%%%%%%%%%%%%%%%

%%%%%%%%%%%%%%%%%%%%%%%%%%%%%%%%%%%%%%%%%%%%%%%%%%%%%%%%%%%%%%%%
%%%%%%%%%%%%%%%%%%%%%%%%%%%%%%%%%%%%%%%%%%%%%%%%%%%%%%%%%%%%%%%%
%%% Missing Data FIGURES
%%%%%%%%%%%%%%%%%%%%%%%%%%%%%%%%%%%%%%%%%%%%%%%%%%%%%%%%%%%%%%%%
%%%%%%%%%%%%%%%%%%%%%%%%%%%%%%%%%%%%%%%%%%%%%%%%%%%%%%%%%%%%%%%%
 \begin{figure*}[ht]
 \hypertarget{Missing data effects} {}
 \subfigure{\includegraphics[width=90mm]{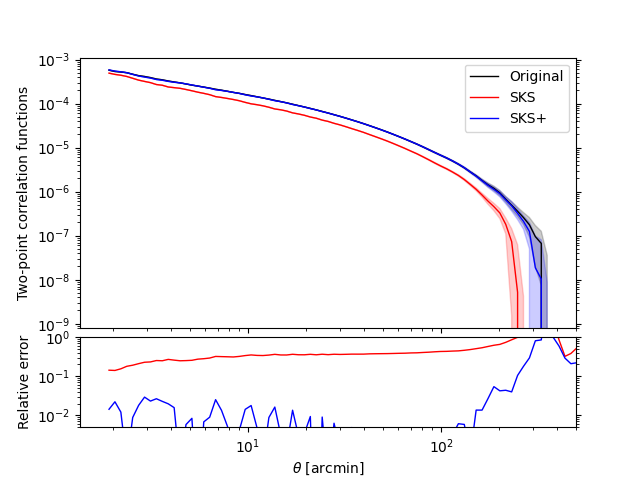}}
 \subfigure{\includegraphics[width=90mm]{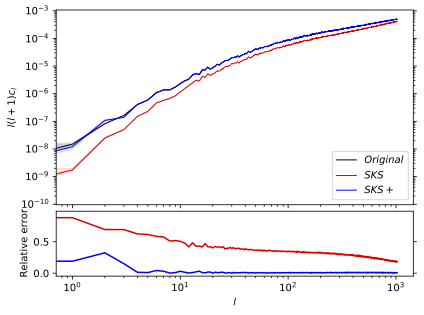}}
 
 \caption{Missing data effects: Mean shear two-point correlation function (black) with corresponding mean convergence two-point correlation function from SKS (red) and SKS+ (blue) reconstruction method (left panel) and power spectrum (right panel) from SKS (red) and SKS+ (blue) reconstruction method, from incomplete noise-free shear maps, compared to the original convergence map (black). The estimation is only made outside the mask $\mathit{M}$. The lower panel shows the relative errors introduced by the missing data effects, that is, the normalised difference between the upper curves.}
 \label{Fig: Missing data}
 \end{figure*}
%%%%%%%%%%%%%%%%%%%%%%%%%%%%%%%%%%%%%%%%%%%%%%%%%%%%%%%%%%%%%%%%
 \begin{figure*}[ht]
 \hypertarget{Missing data Moments} {}
 \subfigure{\includegraphics[width=90mm]{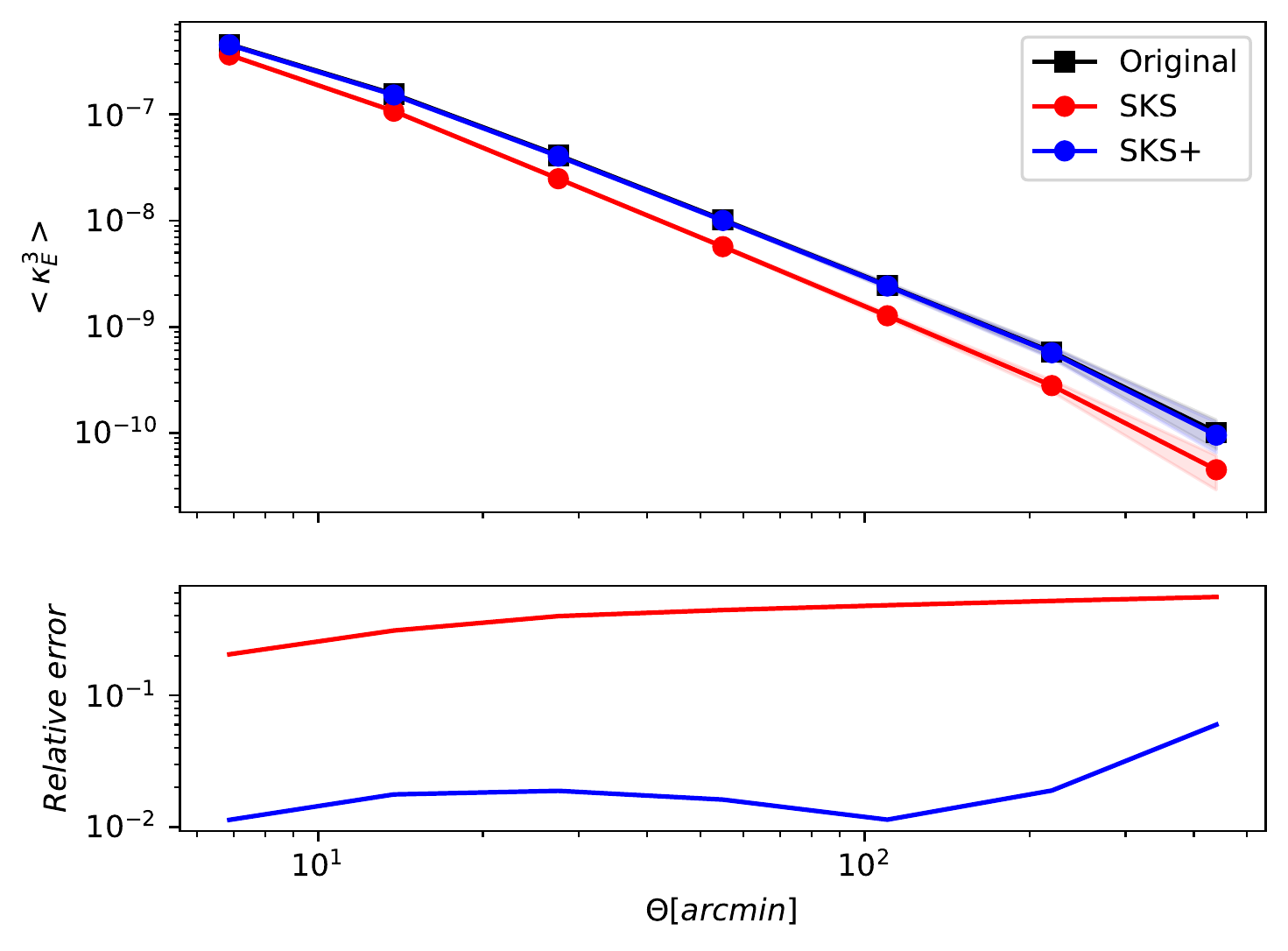}}
 \subfigure{\includegraphics[width=90mm]{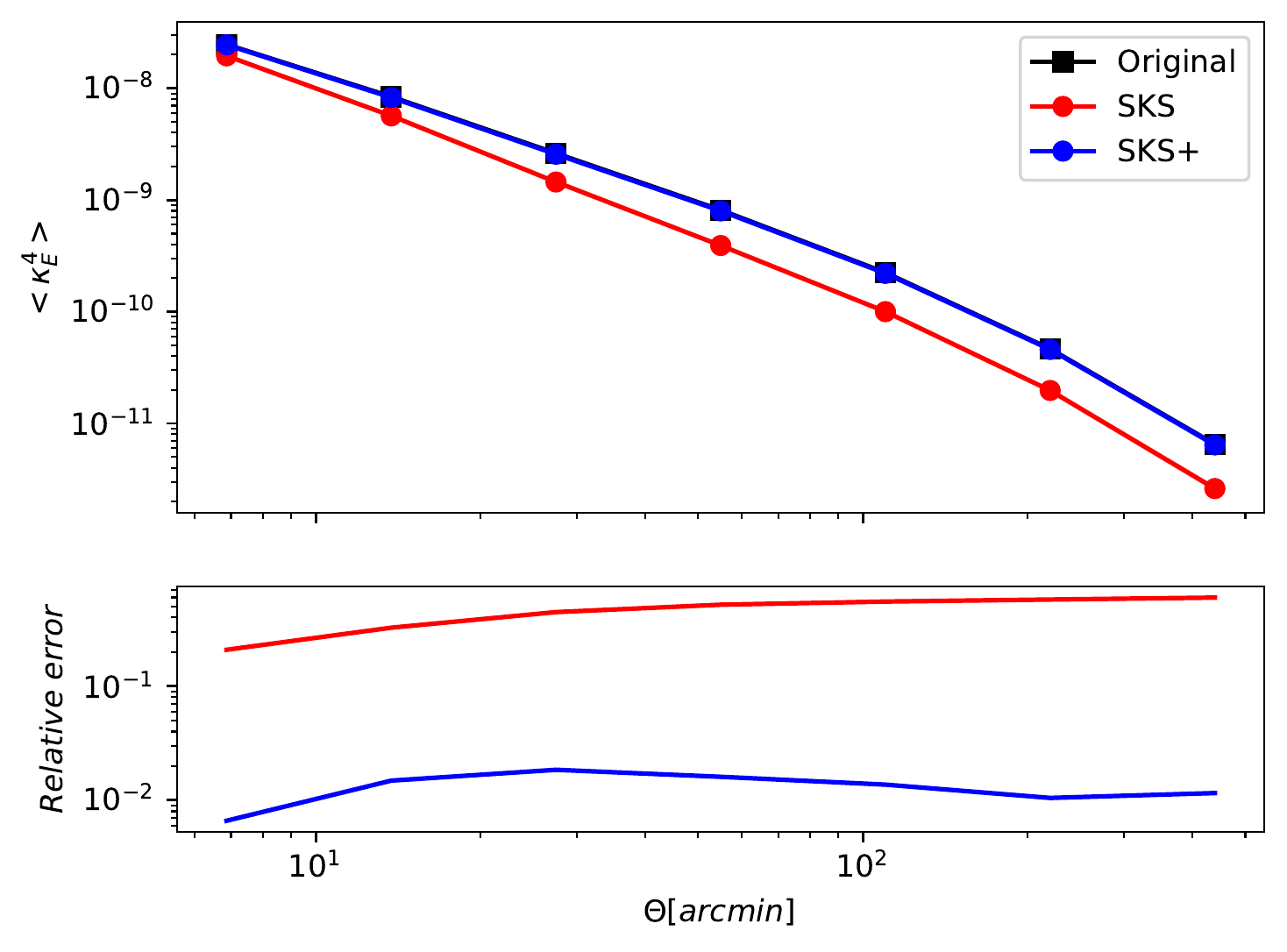}}
 \caption{Missing data effects: third-order (left panel) and fourth-order (right panel) E-mode moments of the original (black) compared to the moments estimated on the SKS (red) and SKS+ (blue) convergence maps on seven wavelet scale. The estimation of the moments is the average of the 10 simulated maps for angular resolution $N_{side}=2048$ that is calculated outside the mask $\mathit{M}$. The lower panel shows the relative higher-order moment errors introduced by missing data effects, that is, the normalised difference between the upper curves.}
 \label{fig:Missing data Moments}
 \end{figure*}
%%%%%%%%%%%%%%%%%%%%%%%%%%%%%%%%%%%%%%%%%%%%%%%%%%%%%%%%%%%%%%
%%%%%%%%%%%%%%%%%%%%%%%%%%%%%%%%%%%%%%%%%%%%%%%%%%%%%%%%%%%%%%%%
\subsection{\textbf{Missing data}}
\label{sec: Missing data}
Masking must be considered when analysing simulated data to more accurately represent real data. Real lensing datasets have gaps due to bad pixels or a defect in the camera's CCD or the presence of a bright star in the field of view, which forces us to remove a part of the image. These gaps are usually described by a simple binary mask. 
These masked areas make post-processing difficult, especially when estimating statistical information such as the power spectrum.

\justify
In this paper, we use the mask that depicts the galactic plane, ecliptic plane, and stellar contamination; it mimics the setting of an upcoming Stage IV survey like Euclid to show how missing data can produce a significant systematic bias on clustering. We also show how our algorithm can accommodate a mask and effectively eliminate any bias. For this purpose, the Euclid-like mask is applied to the corresponding projected noise-free shear map, as shown in the upper panel of the figure \ref{shear-convergence}. The reconstruction of the convergence map is done after masking.
In this section, to demonstrate the effects of the missing data, shape noise effects are not considered. \\
\justify
The middle panel of the second column of figure \ref{shear-convergence} shows the input true convergence map and the first and second columns of the lower panel show the E-mode convergence map reconstructed from the noise-free shear map using SKS and SKS+ method respectively.
\justify
The non-local behaviour of the Kaiser-Squires reconstruction method over limited survey windows introduces masking effects due to unknown shear outside the survey mask in the reconstruction. These masking effects present themselves as additional noise in the reconstructed maps and result in significant leakage (i.e. leakage in the E mode) during mass inversion.
The problem is reformulated taking into account additional assumptions in the SKS+ algorithm, which is defined in \S \ref{sec: Improved Kaiser-Squires}.
\justify
Figure \ref{fig:Residual maps} compares the residual maps (the pixel difference between the input true map and the reconstructed E-mode maps) of the SKS and the SKS+ methods outside the mask. From the residual maps, it can be seen that most of the features of the input convergence map have been recovered, except for the part of the map close to the edges of the mask with the SKS method.
Here we try to quantify the effects of masks arising from the limited survey windows at different scales using the two-point correlation function and higher-order moments.

%%%%%%%%%%%%%%%%%%%%%%%%%%%%%%%%%%%%%%%%%%%%%%%%%%%%%%%%%%%%%%%%
%%% SHAPE NOISE 2PCF FIGURES
%%%%%%%%%%%%%%%%%%%%%%%%%%%%%%%%%%%%%%%%%%%%%%%%%%%%%%%%%%%%%%%%
%%%%%%%%%%%%%%%%%%%%%%%%%%%%%%%%%%%%%%%%%%%%%%%%%%%%%%%%%%%%%%%%

 \begin{figure*}[ht]
 \hypertarget{Shape noise effects} {}
 \subfigure{\includegraphics[width=90mm]{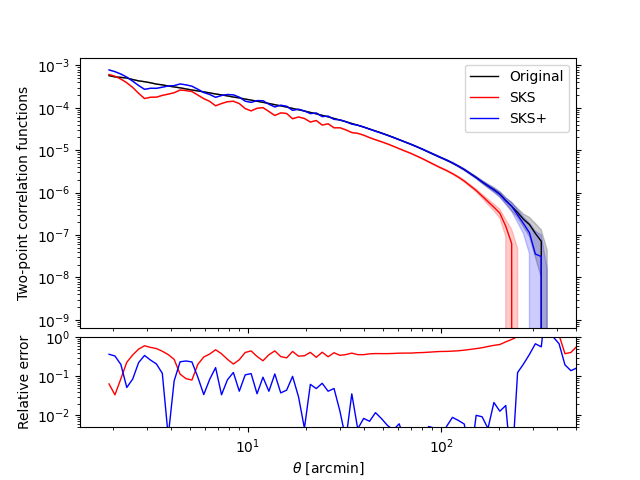}}
 \subfigure{\includegraphics[width=90mm]{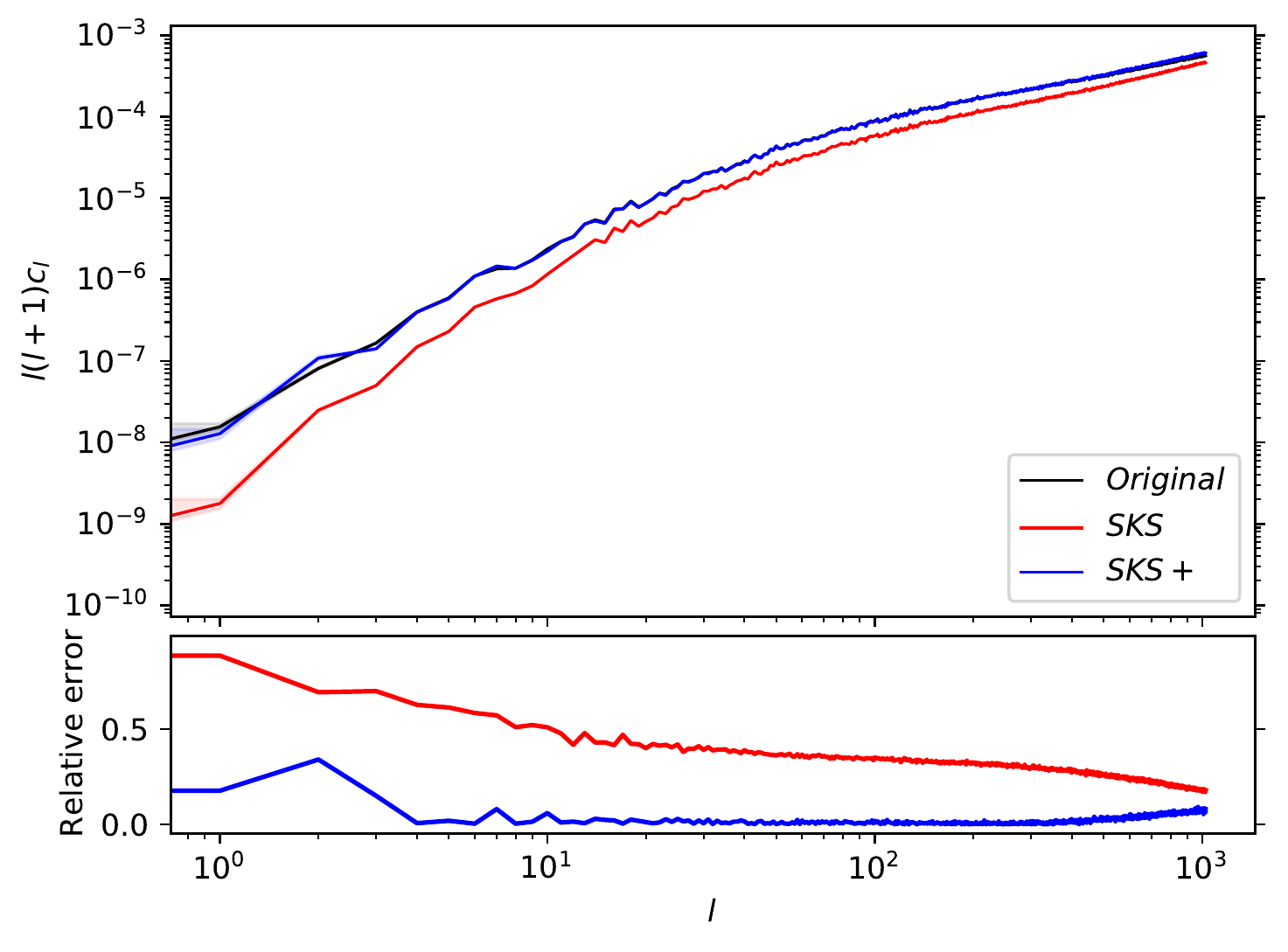}}

 \caption{Shape noise effects: Mean shear two-point correlation function (black) with corresponding mean convergence two-point correlation function from SKS (red) and SKS+ (blue) reconstruction method (left panel) and power spectrum (right panel) from SKS (red) and SKS+ (blue) reconstruction method compared to the original convergence map (black) with realistic shape noise. The estimation is only made outside the mask $\mathit{M}$. The lower panel shows the relative errors introduced by the missing data effects, that is, the normalised difference between the upper curves.}
 \label{Fig: shape noise}
 \end{figure*}

%%%%%%%%%%%%%%%%%%%%%%%%%%%%%%%%%%%%%%%%%%%%%%%%%%%%%%%%%%%%%%%%
%%% SHAPE NOISE Moments FIGURES
%%%%%%%%%%%%%%%%%%%%%%%%%%%%%%%%%%%%%%%%%%%%%%%%%%%%%%%%%%%%%%%%
%%%%%%%%%%%%%%%%%%%%%%%%%%%%%%%%%%%%%%%%%%%%%%%%%%%%%%%%%%%%%%%%
\begin{figure*}[ht]
 \hypertarget{Shape noise Moments} {}
 \subfigure{\includegraphics[width=90mm]{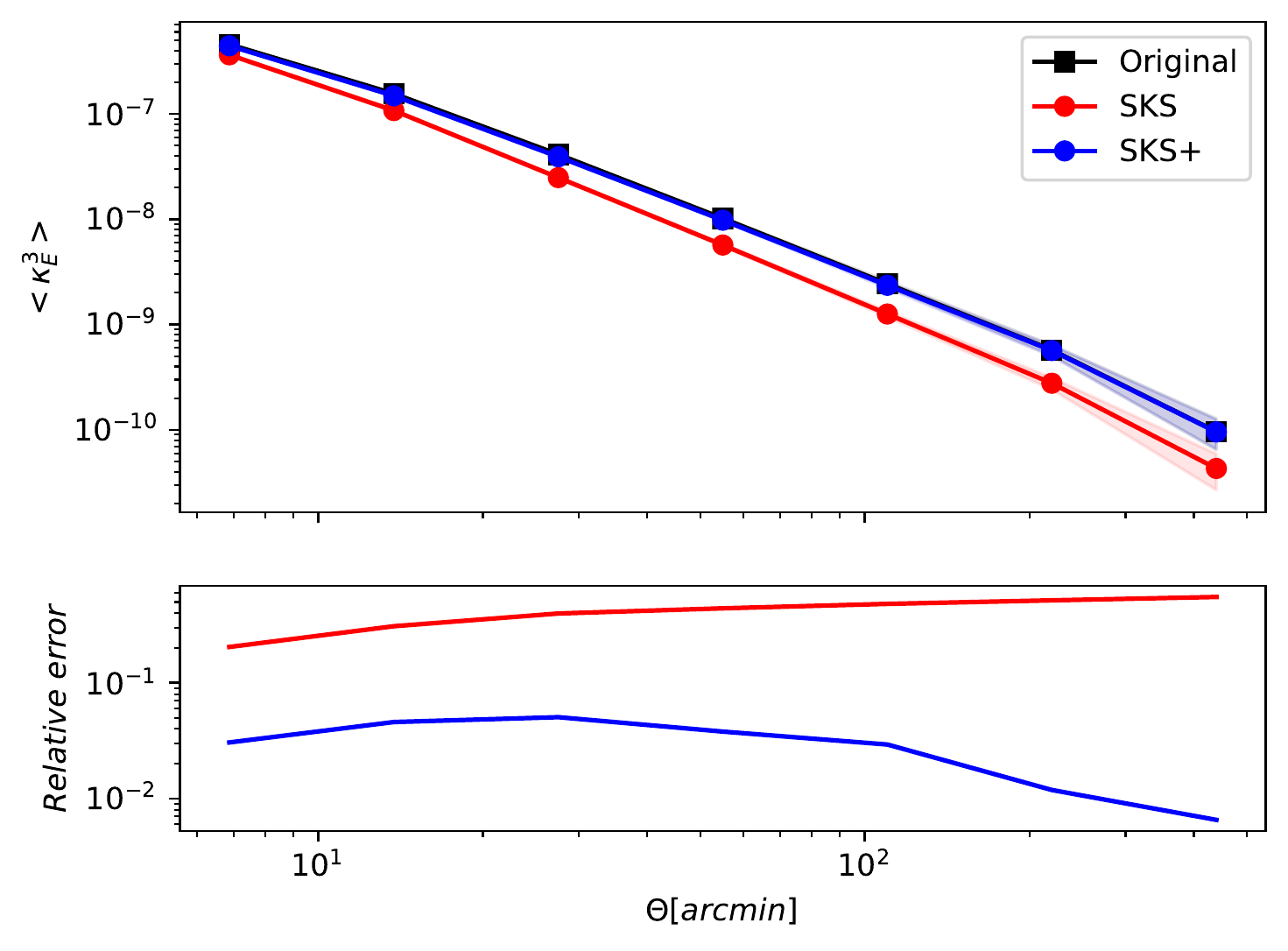}}
 \subfigure{\includegraphics[width=90mm]{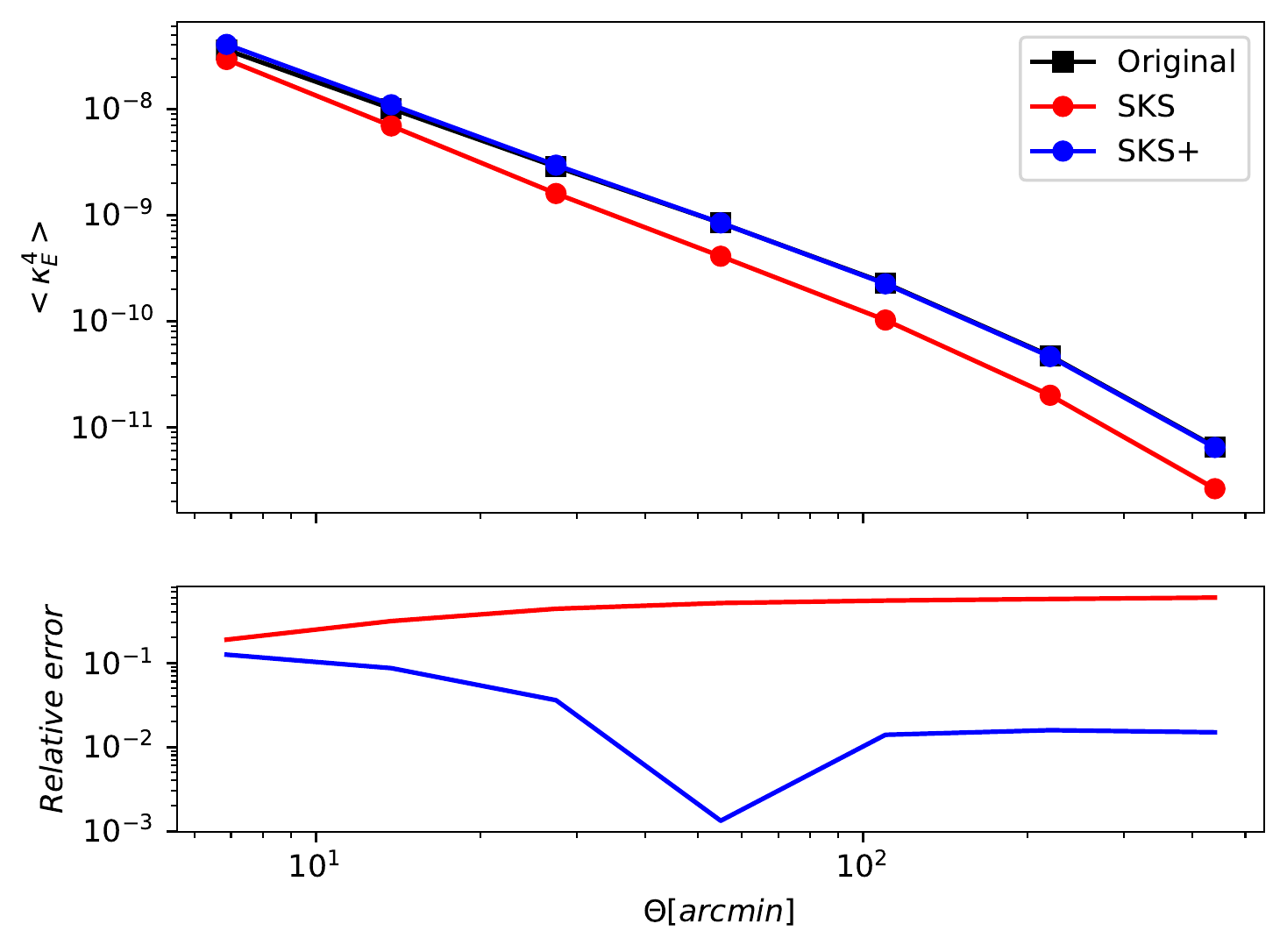}}
 \caption{Shape noise effects: third-order (left panel) and fourth-order (right panel) moments estimated on seven wavelet bands of the original E-mode convergence map with realistic shape noise (black) compared to the moments estimated on the SKS (red) and SKS+ (blue) convergence maps at the same scales. The SKS and SKS+ convergence maps are reconstructed from noisy shear maps. The estimation of the moments is the average of the 10 simulated maps for angular resolution $N_{side}=2048$ that is calculated outside the mask $\mathit{M}$. The lower panel shows the relative higher-order moment errors introduced by shape noise.}
  \label{fig:Shape noise Moments}
 \end{figure*}

%%%%%%%%%%%%%%%%%%%%%%%%%%%%%%%%%%%%%%%%%%%%%%%%%%%%%%%%%%%%%%%%%%%%%%%%%%%
\subsubsection{\textbf{2PCF}}

The two-point correlation function represents the density of two-point distances in a given feature space. The basic weak lensing measurements are the shear two-point correlation function, $\xi_{\pm}$, (2PCF) and its associated power spectrum, the most popular statistic used to capture the second-order statistic in the cosmic shear analyses.
\justify
Shear two-point correlation functions correlate $\gamma_{t/\times}$, the tangential and cross components of shear, of two galaxies separated by an angle $\theta$ in the sky \citep{Bartelmann_2001}.
The two-point correlation function, $\xi_{\pm}$, and the power spectrum, $C_l$, of the weak lensing convergence map $\kappa$ can be used to estimate the shape and normalisation of the dark-matter power spectrum.
The $\xi_{\pm}$ measured with the tree code Athena\footnote{Athena: \url{http://www.cosmostat.org/software/athena}} \citep{Martin_2014}, the estimators defined in \citet{Schneider_2002_2PCF}, with 200 logarithmic angular bins in the range of $1.7 \leq \theta \leq 200$ arcmin. \\
\justify
The angular power spectra, $C_l$, which express the correlations between pairs of galaxy ellipticities in harmonic space, are computed from the original and reconstructed convergence maps using the anafast routine implemented in HEALPix without performing any prior smoothing of the maps. \\

%%%%%%%%%%%%%%%%%%%%%%%%%%%%%%%%%%%%%%%%%%%%%%%%%%%%%%%%%%%%%%%%

\justify
The left panel of figure \ref{Fig: Missing data} compares the mean two-point shear correlation function, $\xi_+$ (black), with the corresponding mean two-point convergence correlation function $\xi_{\kappa_E}$ reconstructed using the SKS method (red) and using the SKS+ method (blue) from incomplete noise-free shear maps. The right panel of the figure \ref{Fig: Missing data} shows the power spectrum estimated from the original convergence map to the one reconstructed from the masked shear maps using the SKS and SKS+ methods. The estimation is done on the 10 simulated maps for angular resolution $N_{side}=2048$ and only made outside the mask $\mathit{M}$. \\
\justify
Figure \ref{Fig: Missing data} clearly show that the two-point correlation function of the SKS reconstruction (red) underestimates the power due to leakage in the shear B modes. The SKS+ map (blue) is a clear improvement over the SKS map and near perfect match to the original two-point correlation function (black).
It is also notable that the SKS method under-predicts the power spectrum at large scales, due to mask effects and E-mode leakage. On the contrary, the SKS+ method recovers the power spectrum more accurately at all scales. The shaded area in both figures represents the error on the mean and the SKS+ falls within the 1$\sigma$ uncertainty from the original estimations. It can be seen that SKS+ has $\sim$ 10 times fewer errors and systematic bias is negligible compared to the SKS method.

%%%%%%%%%%%%%%%%%%%%%%%%%%%%%%%%%%%%%%%%%%%%%%%%%%%%%%%%%%%%%%%%
\subsubsection{\textbf{Moments}}
The moments test is an important and computationally less demanding tool for determining how well the reconstructed convergence maps preserve cosmological characteristics. The skewness of the convergence is sensitive to the cosmological matter density; however, the higher-order moments of the underlying dark matter density fields are much less sensitive to the background cosmological parameters \citep{Jain_2000}. Skewness and kurtosis are also more direct estimators of signal non-Gaussianity.

We performed the wavelet decomposition using \hyperref[sec:UWTS]{UWTS} of our convergence maps, original as well as the reconstructed maps of the mass distribution. For each wavelet scale, we have computed their third-order $\langle \kappa_E^3 \rangle$ and fourth-order $\langle \kappa_E^4 \rangle$ moment outside the masks. 

Figure \ref{fig:Missing data Moments} demonstrates that the SKS+ mass reconstruction method leads to reliable convergence maps with a discrepancy of less than 2\% that preserves the statistical and cosmological information. The SKS errors rise with scale up to a factor of $\sim$3 and are less than 40\% and less than 50\% below 10' and 100', respectively. In contrast, SKS+ errors are substantially less noticeable on all scales and remain within uncertainty $1 \, \sigma$.

%%%%%%%%%%%%%%%%%%%%%%%%%%%%%%%%%%%%%%%%%%%%%%%%%%%%%%%%%%%%%%%%
%%%%% PEak Count Figure
%%%%%%%%%%%%%%%%%%%%%%%%%%%%%%%%%%%%%%%%%%%%%%%%%%%%%%%%%%%%%%%%
 \begin{figure*}[ht]
 \hypertarget{Peak Counts} {}
 \subfigure{\includegraphics[width=90mm]{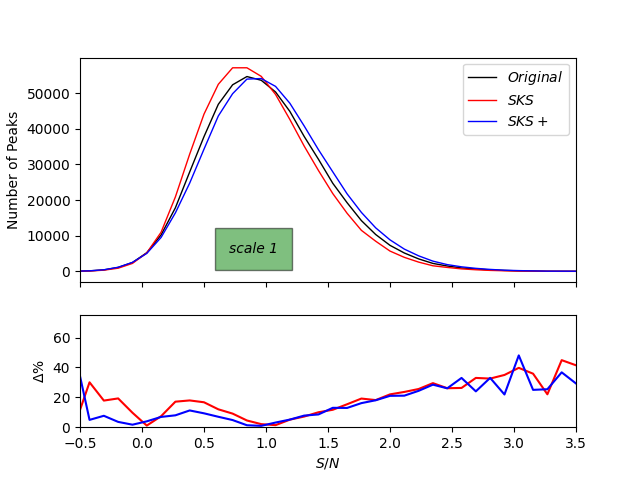}}
 \subfigure{\includegraphics[width=90mm]{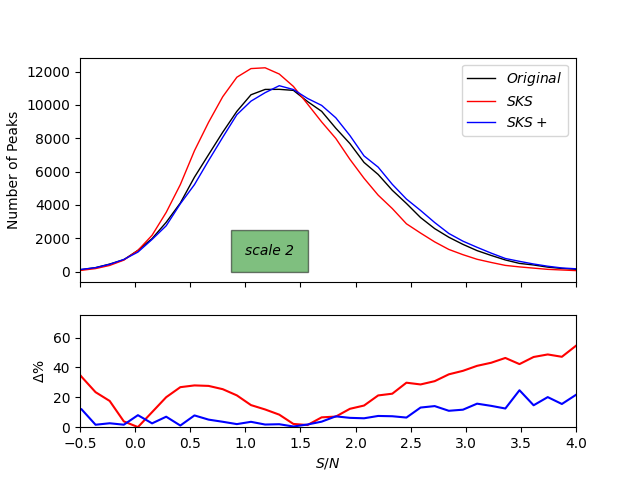}}
  \subfigure{\includegraphics[width=90mm]{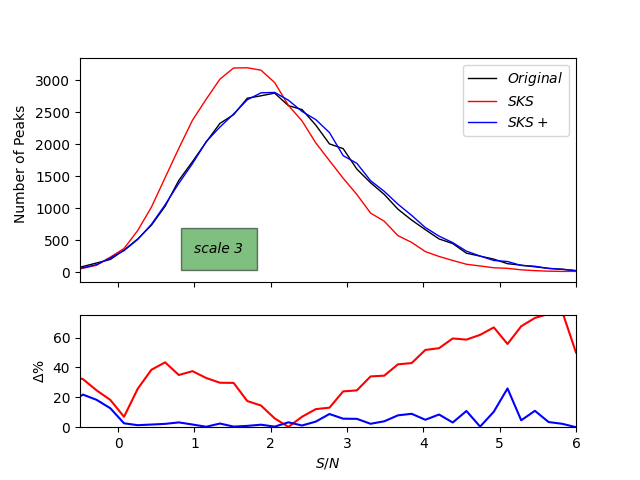}}
 \subfigure{\includegraphics[width=90mm]{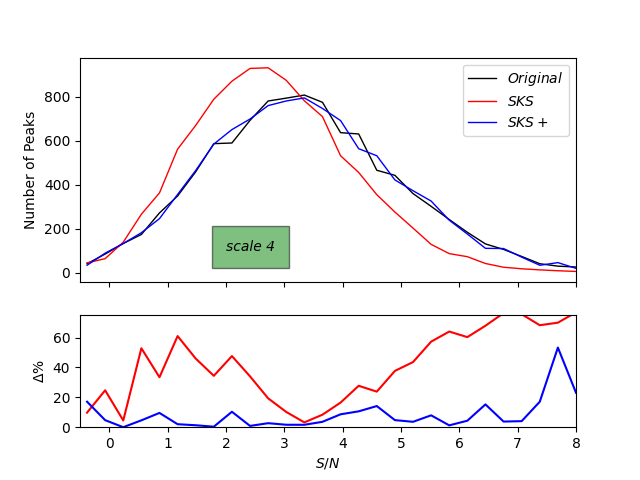}}
 \caption{Peak counts: Convergence map (black) to the Peaks estimated on the SKS (red) and SKS+ (blue) convergence maps over the wavelet coefficient maps \{$w_j$\} up to $j_{max} = 4$. The plots show the average peaks of the 10 simulated maps for angular resolution $N_{side}=2048$ and estimations are only made outside the mask $\mathit{M}$.}
 \label{fig:Peak Counts}
 \end{figure*}
 
%%%%%%%%%%%%%%%%%%%%%%%%%%%%%%%%%%%%%%%%%%%%%%%%%%%%%%%%%%%%%%%%
%%%%%%%%%%%%%%%%%%%%%%%%%%%%%%%%%%%%%%%%%%%%%%%%%%%%%%%%%%%%%%%%
\subsection{\textbf{Weak lensing shape noise}}
\label{sec:Shape noise}
The variance of intrinsic ellipticity, which is the dominant source of noise ($\epsilon_s$), in shear measurements, causes 'shape noise'. The convergence map is then affected by random noise that can be considered Gaussian with zero mean and variance $\sigma_{pix}^2$.
Therefore, we consider the observational errors that are errors from 'shape noise' as expressed in Eq. \eqref{eq19}.
\justify
The noise level of a given pixel in a weak lensing survey is determined by the following factors: the number density of galaxy observations $n_{gal}$ (typically given per $arcmin^{2}$), the size of the said pixel, and the shape noise, the variance of the intrinsic ellipticity distribution, $\sigma_{\epsilon}^{2}$. The variance for an ensemble follows Poisson statistics, so $ \sigma_{pix}^2 = \sigma_{\epsilon}^2 / N$ for an ensemble of N sources, or equivalently, knowing the area $A_{pix}$ of a given pixel with mean source density $n_{g}$, the noise standard deviation $\sigma_{pix}$ is simply given by
\begin{equation} \label{eq26}
 \hypertarget{equation 26} {}
 \begin{aligned}
  \sigma_{pix} = \sqrt{\frac{\sigma_e^{2}}{A_{pix} \times (\frac{180}{\pi})^2 \times 3600 \times n_{gal}}}
 \end{aligned}
\end{equation}\\
\justify
where 3600($\frac{180}{\pi})^{2}$ converts steradians to $arcmin^2$ – this relation is simply a reduction in the noise standard deviation by the root of the number of data points.
As a result, larger pixels that contain more observations (assuming a roughly uniform spatial distribution of galaxy observations) have lower noise variance. In practice, the true number of galaxies in a given pixel, rather than $n_{gal}$, can be used to calculate the value of $\sigma_{pix}$. \\
\justify
In what follows, we assume a typical value of an intrinsic ellipticity standard deviation, $\sigma_{\epsilon}$ $\sim$ 0.26, which is close to typical values for both ground-based and space-based observations \citep[e.g.][]{Leauthaud_2007} and assume a source density, $n_{gal}$ $\sim$ 30 per $arcmin^{2}$, which is roughly what is expected for Euclid \citep{Laureijs_2011}, Roman \citep{Spergel_2015} and LSST \citep{LSST_2009}.\\
\justify
Figure \ref{Fig: shape noise} displays the mean unsmoothed shear two-point correlation function (black) and power spectrum estimates from the unsmoothed convergence maps obtained from SKS and SKS+ methods. The estimation is performed on the 10 simulated maps for angular resolution $N_{side} = 2048$ and only made outside the mask $M$, in the presence of the shape noise. \\

In comparison with the left panel of the figure \ref{Fig: Missing data}, the \ref{Fig: shape noise} shows the less smooth two-point correlation function, caused by the noise fluctuations, which can be reduced simply by increasing the resolution of the HEALPix maps. However, the same findings are obtained that the SKS+ technique has fewer errors than the SKS technique and remains within the 1$\sigma$ uncertainty.\\

Comparing \ref{fig:Missing data Moments} and \ref{fig:Shape noise Moments}, it can be seen that the fourth-order moment is biased at smaller scales. However, the SKS+ method is nearly unbiased and reduces errors by a factor of about 2 and 4 at the third and fourth-order moments, respectively. Furthermore, Figure \ref{fig:Shape noise Moments} represents that the higher-order moments beyond variance, namely skewness and kurtosis, surely preserve the cosmological information even in the presence of a realistic setting.

 %%%%%%%%%%%%%%%%%%%%%%%%%%%%%%%%%%%%%%%%%%%%%%%%%%%%%%%%%%%%%%%%
%%%%% Projection Effects Figure
%%%%%%%%%%%%%%%%%%%%%%%%%%%%%%%%%%%%%%%%%%%%%%%%%%%%%%%%%%%%%%%%
 \begin{figure*}[ht]
 \hypertarget{Projection effects} {}
 \subfigure{\includegraphics[width=90mm]{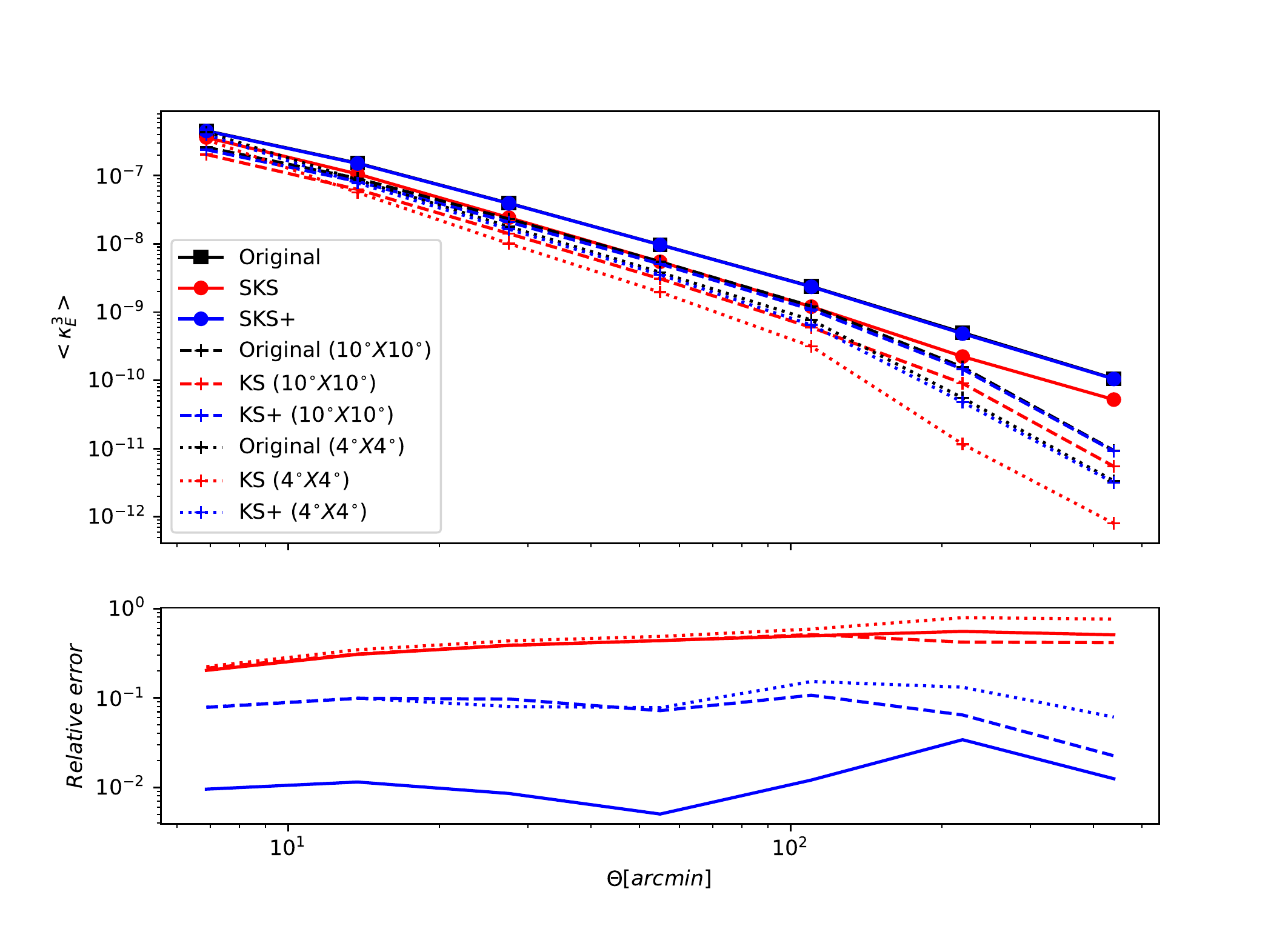}}
 \subfigure{\includegraphics[width=90mm]{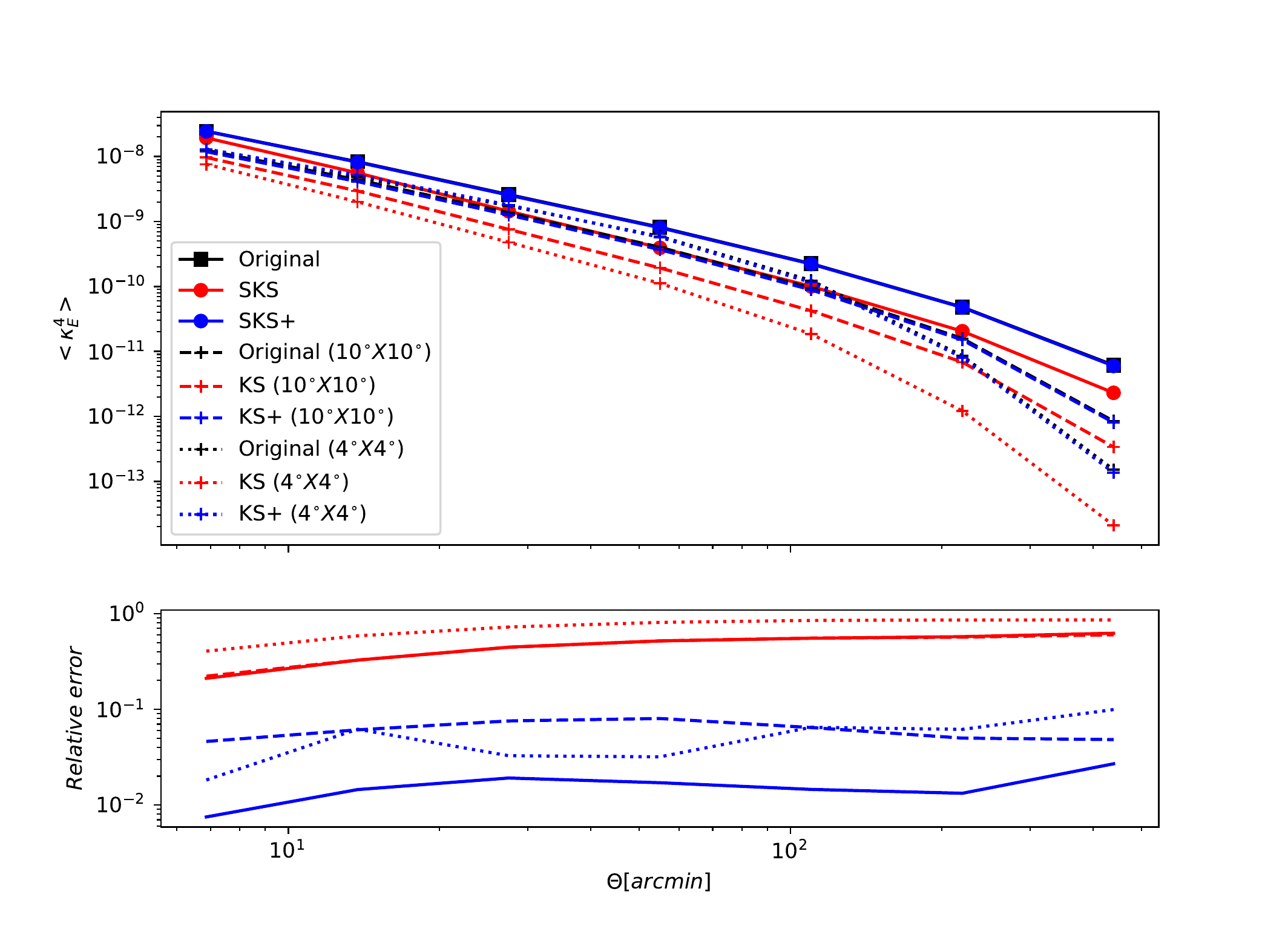}}

 \caption{Projection effects: Third-order (left panel) and fourth-order (right panel) moments estimated on seven wavelet bands of the original E-mode Convergence map (black solid line for full sky, black dashed line and black dotted line for planar approximation of $10^{\circ} \times 10^{\circ}$ and $4^{\circ} \times 4^{\circ}$ respectively) compared to the moments estimated on the SKS, SKS+ (red \& blue solid lines respectively) and KS, KS+ (for planar, red and blue dashed and dotted lines respectively). The plots show the comparison of only one realisation for angular resolution $N_{side}=2048$ and estimations are only made outside the mask $\mathit{M}$.}
 \label{Fig: Pixelisation Effects}
 \end{figure*}
%%%%%%%%%%%%%%%%%%%%%%%%%%%%%%%%%%%%%%%%%%%%%%%%%%%%%%%%%%%%%%%%
%%%%%%%%%%%%%%%%%%%%%%%%%%%%%%%%%%%%%%%%%%%%%%%%%%%%%%%%%%%%%%%%
%%%%%%%%%%%%%%%%%%%%%%%%%%%%%%%%%%%%%%%%%%%%%%%%%%%%%%%%%%%%%%%%
\subsection{\textbf{Peak counts}}
\label{sec: Peak counts}

Peak counts are a simple and robust statistic that has received increasing attention in recent years \citep{ Maturi_2010, Yang_2011, Lin_2015, Peel_2018, Ajani_2020, Emma_2022}, for accessing cosmological information from the non-Gaussianities in lensing fields.
The peaks can be used as tracers of dark-matter halos, galaxy clusters, and lensing structure, and therefore the non-Gaussianities in the matter distribution. Therefore, the peaks are sensitive to the halo mass function and cosmological parameters \citep{Marian_2009, shirasaki_2017}. In this work, the focus is only on the study of peak counts, and the exploration of peak profiles and peak-correlation functions to improve cosmological constraints is left for further studies.\\

The peak count statistic is a straightforward evaluation that determines the number of peaks on the reconstructed convergence map as a function of their signal-to-noise ratio (SNR). Peak counting is accomplished by scanning pixels in a signal-to-noise field $S/N$, and looking for local maxima (i.e., pixels with a higher value of $\kappa$ than its surrounding 8 pixels). A pixel is regarded as a peak if its value is higher than all the values of its neighbouring pixels.
The $get\_all\_neighbours$ function is used to search all neighbouring pixels. The number of peaks is then counted as a function of their centre value. 
Where $S/N$ on the scale $j$ is the ratio between the convergence $\kappa_j$ and the standard deviation of the noise.

\begin{equation} \label{PC_w}
 \begin{aligned}
  (S/N)_j = \frac{\kappa_j}{\sigma_{noise}}
 \end{aligned}
\end{equation}

\justify
In this case, the estimation of the standard deviation of the noise, $\sigma_{noise}$, is performed on each wavelet scale, as described in \citet{Starck_1998}. The $\kappa_j$ is the set of aperture mass maps that are obtained by convolving $\kappa$ with 1D filter $h_j$ as shown in Eq. \eqref{eq12}. We transform $\kappa$ with the wavelet scale $J = 4$ and identify the peaks on these maps. 
\justify 
The features of the map have been grouped into bins of the signal-to-noise ratio (SNR) instead of $\kappa$, as has been done mainly in previous studies. The positive and negative peaks of the full sky convergence maps without tomographic decomposition are estimated outside the mask. Figure \ref{fig:Peak Counts} shows the peak abundance as a function of the SNR. Peaks estimated on original and reconstructed convergence maps are shown in black, red (SKS) and blue (SKS+) respectively. The curves correspond to the mean over 10 simulated maps for angular resolution $N_{side} = 2048$. The SKS+ method provides on average 20-30\% better peak estimation compared to the SKS method.
It can be seen that the number of counts depends on the resolution, as expected, and the transform picks out features of the convergence map, $\kappa$, at successively larger scales as $j$ increases. Figure \ref{fig:Peak Counts} also shows the good agreement between the SKS+ and the original predictions on all wavelet filtering scales. It is also noteworthy that peak statistics are likely to be robust towards unknown errors in the galaxy ellipticities that complicate the use of field statistics.
In principle, peak statistics encode information about dark matter halos, voids, and cluster structures, and therefore non-Gaussianities in matter distribution \citep{fan_2010, shirasaki_2017}. Further work is required to evaluate the biasing of galaxies relative to dark matter by estimating the relation of galaxies to halos and the relation of halos to dark matter.

%%%%%%%%%%%%%%%%%%%%%%%%%%%%%%%%%%%%%%%%%%%%%%%%%%%%%%%%%%%%%%%%
\subsection{\textbf{Projection effects}}

As moving towards greater sky-coverage areas with near-future surveys, namely Euclid (15,000 $deg^2$) and LSST (20,000 $deg^2$) compared to DES (5000 $deg^2$), it is necessary to understand the projection effects, take sky geometry into account, and encourage spherical analysis over planar approximation. Spherical projections onto the plane cannot preserve all the features of the spherical map, and these projection effects decrease the sensitivity to the clustering on small scales. \\

This section aims to understand the effects of projections; therefore, only noiseless maps are considered. Mapping the spherical projection from HEALPix to the plane or vice-versa results in distortion. 
When creating convergence maps on the plane, the exact projection used to map the celestial sphere to the plane can have an impact on the quality of the reconstructed convergence map. Since the resulting spherical harmonic transforms are theoretically exact, any errors will therefore be due to projection effects rather than inaccuracies in the harmonic transforms. \\

We compared the higher-order moment, estimates in the E-mode of the spherical convergence map and planar convergence maps as in figure \ref{Fig: Pixelisation Effects}.
To investigate the distortions, there was an absolute need to partition the full-sky shear maps, with and without masks, into patches and, preferably, non-overlapping planar maps. In practice, the size of the patches plays an important role in obtaining noticeable projection effects.
Therefore, we decompose the celestial sphere into square patches of $4 \times 4 \, deg^2$ and $10 \times 10 \, deg^2$ (centered on the centre of the planar map) in such a way that the induced distortions remain at a minimum and the pixel size corresponds to $N_{side}$ = 2048. The projected HEALPix pixels of the flat maps are converted into standard square pixels using nearest-neighbour interpolation. \\

The convergence field was estimated from this planar shear field using the planar KS \citep{KS_1993} and KS+ \citep{Pires_2020} estimators, and then compared these recovered planar convergences to the celestial sphere convergence field estimated using SKS (\S \ref{sec: Kaiser-Squires on the sphere}) and SKS+ (\S \ref{sec: Improved Kaiser-Squires}) methods. \\

As shown in figure \ref{Fig: Pixelisation Effects}, the pixelation effect biases the moments too much which is minimal for small and intermediate scales but more significant for the largest scales. This bias can be explained due to the projection distortion and the boundary created by the projection.
It can be seen that convergence reconstruction on the celestial sphere not only reduces errors but is also easily generalised by replacing the usual flat-sky convergence reconstruction. However, due to the non-linear nature of the effect, induced by the projection remain at a 10\% level on smaller scales. \\

We also test the projection effects by converting HEALPix pixels into standard square pixels using bi-linear interpolation and we didn't find any significant difference.\\
Instead of just decomposing shear maps and performing KS and KS+, we also decomposed our convergence maps reconstructed from SKS and SKS+ methods, and we got the equivalent yet somewhat better results at small scales. Hence, we can say that by recovering weak lensing mass maps directly on the sphere and thereby eliminating the distortion due to a projection to the plane, the performance of reconstruction is enhanced. Reconstruction of the convergence field directly on the celestial sphere also allows us to consider or exclude the specific regions of the sphere.  

%%%%%%%%%%%%%%%%%%%%%%%%%%%%%%%%%%%%%%%%%%%%%%%%%%%%%%%%%%%%%%%%
\section{Conclusions}
\label{sec: Summary and Conclusion}
In this work, a new method, SKS+, for reconstructing mass maps without noise regularisation from weak lensing shear observations has been discussed. The previously presented KS+ \citep{Pires_2020} reconstruction formalism is extended to the spherical setting, resulting in a spherical KS+ mass-mapping algorithm which refers to as the SKS+ method. Throughout the paper, the analysis is performed at a typically adopted angular band limit of $l_{max} = 3 * N_{side} - 1$ and $N_{side} = 2048$. A higher resolution can result in extremely high computational costs, whereas a lower resolution would destroy the data. SKS+ method, on the other hand, is applicable to any pixelisation of the sphere. \\

\S \ref{sec: Spherical Mass mapping reconstruction} described how one can recover convergence fields directly on the celestial sphere by adopting the SKS and SKS+ techniques. The effects of missing data, sampling, and shape noise were tested and compared on the SKS and SKS+ reconstructed convergence maps using two-point correlation, power spectrum, third-and fourth-order moments, and peaks in a realistic setting.
It is shown that the SKS+ method outperforms and reduces errors compared to the SKS method and preserves the statistical and cosmological information. In the noise-free case, we show that the SKS+ method systematically improves the inferred correlation errors that are affected by the Euclidean mask $\sim$ 10 times compared to the SKS method. We also showed that the SKS method systematically underestimates the higher-order moments i.e. skewness and kurtosis at all scales and the SKS+ method is nearly unbiased. While after adding realistic noise, the SKS+ method reduces errors by a factor of about 2 and 4 in the third- and fourth-order moments, respectively. These measurements from convergence maps may provide important constraints on the cosmological parameters $\Omega_m$ and $\sigma_8$. Furthermore, the results showed that the SKS+ reconstructed convergence maps using N-body simulations \citep{Takashi_2017} are unbiased in relation to truth, which is properly negligible.
The SKS+ method has been shown to provide on average 20-30\% better maximum signal-to-noise peak estimations compared to the SKS method, which can be a useful new approach to wide-field lensing surveys to encode information about dark-matter halos, voids, and cluster structures, and a useful test of the presence of systematic errors. \\
\justify 
The SKS and SKS+ methods reduce to the usual flat-sky approach KS \citep{KS_1993} and KS+ \citep{Pires_2020} approach in the planar approximation. The accuracy of the planar approximation for mass mapping has been studied and addresses the important question of whether one needs to recover the convergence field on the sphere for forthcoming surveys or whether recovery on the plane would be sufficient. It has been found that projection errors remain at a 10\% level on small scales, which can decrease the sensitivity to the clustering. These projection errors can be eliminated by recovering convergence maps directly on the celestial sphere by the SKS and SKS+ techniques. Surely, one can decompose the spherical convergence map into $1^\circ \times 1^\circ$ patches and include or exclude parts of the sky, according to the needs of the processing. We believe that the solution is more viable to reduce storage costs than using the planar mass inversion.\\
\justify
While planar projection analysis is less computationally demanding than sphere analysis, the transition in research towards analysing full-sky is necessary to derive and compare statistics containing cosmological information to the flat-sky approximation. It will be of important use for application to Stage IV datasets like Euclid, Roman, and LSST to avoid the significant errors that are induced by planar approximations. \\
\justify
Finally, the SKS+ method provides dramatically increased reconstruction fidelity over the SKS method. SKS+ method is successful at capturing the non-Gaussianity that provides direct detection of non-linear evolution of the dark matter. Furthermore, the SKS+ method is easily generalised to the full sky by replacing KS+. 
Moreover, the SKS+ method has the ability to preserve the sharpness of clusters and the gravitational nature of the signal, which could be interesting in the estimation of the cosmological parameters, the physics of clusters, the profiles around peaks, peak-correlation functions, etc. 
Future work using this method will address improving cosmological constraints, constructing a mass-selected halo catalogue \& measuring its statistical properties, and evaluating the biasing of the galaxies relative to the dark matter using peak counts and higher-order moments.

%%%%%%%%%%%%%%%%%%%%%%%%%%%%%%%%%%%%%%%%%%%%%%%%%%%%%%%%%%%%%%%%
%%% Acknowledgments
%%%%%%%%%%%%%%%%%%%%%%%%%%%%%%%%%%%%%%%%%%%%%%%%%%%%%%%%%%%%%%%%

\begin{acknowledgements}
The author thanks Ma'am Sandrine Pires for the discussions. I would like to thank Euclid Consortium for allowing me to use their resources.
I am very grateful to EN Taylor, Henk Hoekstra and Koen Kuijken for helpful, illuminating and useful comments on a pre-submission draft. I am also grateful to Nicolas Laporte for his help.

\end{acknowledgements}

%%%%%%%%%%%%%%%%%%%%%%%%%%%%%%%%%%%%%%%%%%%%%%%%%%%%%%%%%%%%%%%%
\nocite{*}
%%%%%%%%%%%%%%%%%%%%%%%%%%%%%%%%%%%%%%%%%%%%%%%%%%%%%%%%%%%%%%%%
% To change the title from References to Bibliography:
\renewcommand\refname{Bibliography}
\bibliographystyle{aa} % style aa.bst
\bibliography{aa} % your references Yourfile.bib

\end{document}